\documentclass[12pt]{article}
\usepackage{amssymb}
\usepackage{amsmath}
\include{diagxy}

\begin{document}
%==================================================================
\newtheorem {proposition}{Proposition}[section]
\newtheorem{lemma}{Lemma}[section]
\newtheorem{theorem}{Theorem}[section]
\newtheorem{corollary}{Corollary}[section]
\newtheorem{remark}{Remark}[section]

%\numberwithin{equation}{section}

\numberwithin{equation}{section}

\bigskip
\bigskip
\begin{center}
{\large\bf POSITIVE KERNELS AND QUANTIZATION}
\end{center}
\bigskip
\bigskip
\begin{center}
{\bf Anatol Odzijewicz\footnote{ E-mail: aodzijew@uwb.edu.pl},
Maciej Horowski}\footnote{ E-mail: horowski@math.uwb.edu.pl}

\end{center}
\bigskip
\bigskip
\begin{center}
{Institute of Mathematics\\ University in Bia{\l}ystok
\\Lipowa 41, 15-424 Bia{\l}ystok, Poland}
\end{center}
\bigskip\bigskip

\begin{abstract}
\noindent In the paper we investigate a method of quantization based
on the concept of positive definite kernel on a principal $G$-bundle
with compact structural group $G$. For $G=U(1)$ our approach leads
to Kostant--Souriau geometric quantization as well as to coherent
state method of quantization. So, the theory proposed here can be
treated as a generalization of both mentioned quantizations to the
case of general compact group.
\end{abstract}
\vspace{1.5cm} \thispagestyle{empty}

\tableofcontents

\section{Introduction}

The geometric quantization initiated in \cite{K,Kir,Sou} provides an
effective machinery for quantization of Hamiltonian systems. In fact
the main ingredient of Kostant-Souriau theory is a principal
$U(1)$-bundle $\pi:P\to M$ over a symplectic manifold $(M,\omega)$
with connection form $\vartheta$ which satisfies the consistency
condition $\pi^*\omega=i\; curv\, \vartheta$ with the symplectic
form $\omega$.

The other essential element of this theory is a complex distribution
$\mathcal P\subset T^{\mathbb C}M$ which is maximal and isotropic
with respect to $\omega$, called the polarization. The crucial step
for quantization of a physical classical observable $f\in
C^\infty(M)$ is the proper choice of polarization $\mathcal P$. Then
one constructs for $f$ a selfadjoint operator $\hat F$ (quantum
observable) which acts in the Hilbert space consisting of such
sections of the corresponding prequantum bundle which are
annihilated by $\mathcal P$, e.g. see \cite{S}.

The coherent state method of quantization is based on the concept of
coherent state map, i.e. a symplectic map $\mathcal K$ of the phase
space $(M,\omega)$ into the quantum phase space of pure states
$(\mathbb{CP}(\mathcal H),\omega_{FS})$ which is the complex
projective Hilbert space with Fubini--Study form $\omega_{FS}$ as
the symplectic form. The K\"ahler form $\omega_{FS}$ is the
curvature form of the connection form $\vartheta_{FS}$ defined
canonically on the tautological principal $U(1)$-bundle
$\pi:P_{FS}\to\mathbb{CP}(\mathcal H)$ by the metric and complex
structure of $\mathcal H$.

Moreover the scalar product in Hilbert space $\mathcal H$ also
defines the positive definite kernel $K_{FS}:P_{FS}\times P_{FS}\to
\mathbb C$, which after normalization has  a physical  interpretation as
the transition amplitude between two pure states. The canonical
prequantum $U(1)$-bundle $(\pi:P_{FS}\to\mathbb{CP}(\mathcal
H),\vartheta_{FS},\omega_{FS})$ as well as
$(\pi:P_{FS}\to\mathbb{CP}(\mathcal H),K_{FS})$ are the universal
objects in the category of all prequantum $U(1)$-bundles and in the
category of principal bundles with fixed positive definite kernels
$(\pi:P\to M,K)$, respectively. Between these categories there is
functorial dependence, see \cite{O1}, i.e. any prequantum bundle
$(\pi:P\to M,\vartheta,\omega)$ is obtained from $(\pi:P\to M,K)$
for some properly chosen kernel $K$. In \cite{O1} also it is shown
that one can quantize those Hamiltonian flows on $(M,\omega)$ which
preserve the kernel $K$.

Motivated by the fundamental role of positive definite kernels in
the geometry of the prequantum bundles as well as their physical
interpretation as transition amplitudes we investigate here method
of quantization entirely based on the notion of this type of kernels
in the case of general compact structural group $G$ (for $G=U(1)$
see \cite{O1}). In fact we quantize the one-parameter groups
$\{\tau_t\}_{t\in\mathbb R}$ of automorphisms of the principal
$G$-bundles with fixed positive definite kernels $(\pi:P\to M,K)$.
The projection $\sigma_t(\pi(p)):=\pi(\tau_t(p))$, $p\in P$, of the
flow $\{\tau_t\}_{t\in\mathbb R}$, on $M$ is described by the
equation \eqref{h71am87i} which  is a version of the Hamiltonian
equation for the flow $\{\sigma_t\}_{t\in\mathbb R}$. This equation
connect the vector field $X$ tangent to $\{\sigma_t\}_{t\in\mathbb
R}$ (Hamiltonian vector field) with generating function
$F:P\to\mathcal B(V)$ which is $G$-equivariant and operator valued
(Hamiltonian). See \eqref{h71am87i} and Proposition \ref{prop:3.1}.

This paper is organized as follows. In Section 2 we give a short
outline of the theory of positive definite kernels. Especially we
investigate these kernels on the principal G-bundles describing
their relationship with the notion of connection.

In Section 3 and Section 4 we show that for the pair $(F,X)$,
satisfying (\ref{h71am87i}), one can construct the $G$-version of
Kostant--Souriau operator $Q_{(F,X)}$, see \eqref{osi4e} and
\eqref{pre32q1} or \eqref{pre32q13} in non-singular case. The
differential operator $Q_{(F,X)}$ can be extended to a self-adjoint
operator in Hilbert space completely defined by the positive
definite kernel $K:P\times P\to \mathcal B(V)$ (see Theorem
\ref{th:4.1} and Proposition \ref{prop:4.2}).

Finally, we present two simple examples illustrating the method of
quantization proposed here. One can find other examples important
for physical applications in \cite{O-S,H-O,H-O-T}.

\section {Positive definite kernels on  principal bundles}

We begin this section with a short presentation of the theory of
operator-valued positive definite kernels. A more exhaustive
treatment can be found for example in Chapter I of \cite{N}.

Let us take a set $P$ and complex Hilbert spaces $V$ and ${\cal H}$.
By $\mathcal{B}(V,{\cal H})$ we denote the Banach space of bounded
linear operators from $V$ into ${\cal H}$. By $\mathcal{B}(V)$ we
denote $\mathcal{B}(V,V)$. For adjoint of $A\in\mathcal{B}(V,{\cal
H})$ we will write $A^*\in\mathcal{B}({\cal H},V)$.

Now we will show that there exist functorial correspondences between
three categories whose objects are the following:

\textbf{(i)} $\mathcal{B}(V)$-valued positive definite kernels, i.e.
maps $K:P\times P\rightarrow\mathcal{B}(V)$ such that for any finite
sequences $p_1,\ldots,p_J\in P$ and $v_1,\ldots,v_J\in V$ one has
\begin{equation}\label{b5iu}
\sum_{i,j=1}^J \langle v_i, K(p_i,p_j)v_j\rangle\geqq 0,
\end{equation}
where $\langle\cdot,\cdot\rangle$ denotes the scalar product in $V$.
Everywhere in the paper we assume that scalar products are
anti-linear in the first argument and linear in the second argument.

The positivity condition (\ref{b5iu}) implies that $K$ is
Hermitian, i.e. for each $q,p\in P$ one has
\begin{equation}\label{b4iu}
K(q,p)=K(p,q)^*.
\end{equation}

\textbf{(ii)} maps $\mathfrak{K}:P\rightarrow \mathcal{B}(V,{\cal
H})$ satisfying the condition
\begin{equation}\label{gyt5}
    \{\mathfrak{K}(p)v: p\in P \;\;and\;\; v\in V \}^\perp=\{0\}.
\end{equation}

\textbf{(iii)} Hilbert spaces ${\cal K}\subset V^P$ realized by
V-valued functions $f:P\rightarrow V$ such that the evaluation
functionals
\begin{equation}\label{ewa7}
    E_pf:=f(p)
\end{equation}
are continuous maps of Hilbert spaces $E_p:{\cal K}\rightarrow V$
for every $p\in P$.

These functorial correspondences for the case $G=U(1)$ and
$\dim_{\mathbb{C}}V=1$ are proved in \cite{O1}. The proofs of these
correspondences for the case of general group $G$ and general
Hilbert space $V$ can be given in a similar way. Here we restrict
our considerations to the main steps of these proofs.

Equivalence between \textbf{(ii)} and \textbf{(iii)} is given as
follows. For $\mathfrak{K}:P\rightarrow \mathcal{B}(V,{\cal H})$ we
define monomorphism of vector spaces $J:{\cal H}\rightarrow V^P$ by
\begin{equation}\label{ne3e}
    J(\psi)(p):=\mathfrak{K}(p)^*\psi,
\end{equation}
where $\psi\in{\cal H}$. Using this monomorphism we obtain Hilbert
space structure on ${\cal K}:=J({\cal H})$. The continuity of the
evaluation functionals follows from the inequality
\begin{equation}\label{cewa2}
    \|E_pJ(\psi)\|=\|\mathfrak{K}(p)^*\psi\|\leq
    \|\mathfrak{K}(p)^*\|\cdot\|\psi\|=\|\mathfrak{K}(p)^*\|\cdot\|J(\psi)\|.
\end{equation}
Taking Hilbert space ${\cal K}\subset V^P$ such as in \textbf{(iii)}
we put by definition ${\cal H}:={\cal K}$ and define
$\mathfrak{K}(p):V\rightarrow{\cal H}$ by
\begin{equation}\label{kr4e}
\mathfrak{K}(p):=E_p^*.
\end{equation}
In order to check (\ref{gyt5}) note that
\begin{equation}\label{drr4e}
\langle\mathfrak{K}(p)v \mid f\rangle=\langle E_p^*v \mid f\rangle=
\langle v,E_pf\rangle=\langle v,f(p)\rangle
\end{equation}
where $\langle\cdot|\cdot\rangle$ is the scalar product in ${\cal
H}$. Thus if $\langle\mathfrak{K}(p)v | f\rangle=0$ for every $v\in
V$ and $p\in P$, then \eqref{gyt5} implies $f(p)=0$ for all $p\in
P$.

Since one has
\begin{equation}\label{je32e}
0\leq\left\|\sum_{i=1}^J\mathfrak{K}(p_i)v_i  \right\|^2
=\sum_{i,j=1}^J\langle
v_i,\mathfrak{K}(p_i)^*\mathfrak{K}(p_j)v_j\rangle,
\end{equation}
the passage from \textbf{(ii)} to \textbf{(i)} is given by
\begin{equation}\label{f4r54}
    K(q,p):=\mathfrak{K}(q)^*\mathfrak{K}(p).
\end{equation}

In order to show the implication \textbf{(i)} $\Rightarrow$
\textbf{(iii)} let us take vector subspace ${\cal K}_0\subset V^P$
consisting of functions
\begin{equation}\label{Afr432}
    f(p):=\sum_{i=1}^I K(p,p_i)v_i,
\end{equation}
defined for finite sequences $p_1,\ldots,p_I\in P$ and
$v_1,\ldots,v_I\in V$. Due to positive definiteness of the kernel
$K:P\times P\rightarrow\mathcal{B}(V)$ we can define a scalar
product between $g(\cdot)=\sum_{j=1}^{J}K(\cdot,q_j)w_j\in{\cal
K}_0$ and $f\in{\cal K}_0$ by the formula
\begin{equation}\label{sc3e}
    \langle g|f\rangle:=\sum_{i=1}^I\sum_{j=1}^{J}\langle
    K(p_i,q_j)w_j,v_i\rangle.
\end{equation}
Substituting $g(\cdot)=K(\cdot,p)v\in{\cal K}_0$ into (\ref{sc3e})
we obtain the reproducing property
\begin{equation}\label{r6e43}
    \langle v,f(p)\rangle=\langle
    v,\sum_{i=1}^I K(p,p_i)v_i\rangle= \sum_{i=1}^I \langle
    K(p_i,p)v,v_i\rangle =\langle K(\cdot,p)v\mid f\rangle.
\end{equation}
From (\ref{r6e43}) one has the inequality
\begin{equation}\label{f443q}
    \|f(p)\|\leq\sqrt{\|K(p,p)\|}\|f\|,
\end{equation}
which proves that (\ref{sc3e}) is a positive definite scalar
product. Inequality (\ref{f443q}) implies that if $\{f_n\}$ is a
fundamental sequence in ${\cal K}_0$, then $\{f_n(p)\}$ is a
fundamental sequence in $V$. Thus one can realize equivalence
classes of fundamental sequences $[\{f_n\}]\in\overline{{\cal K}_0}$
by the function
\begin{equation}\label{je332}
    f(p):=\lim_{n\rightarrow\infty}f_n(p)
\end{equation}
from $V^P$. In consequence we embed $\iota:\overline{{\cal
K}_0}\hookrightarrow V^P$ as the abstract complement $\overline{{\cal
K}_0}$ of ${\cal K}_0$ into $V^P$. Summing up we define Hilbert
space ${\cal K}\subset V^P$ from \textbf{(iii)} as ${\cal
K}:=\iota(\overline{{\cal K}_0})$. The continuity of the evaluation
functionals $E_p$, $p\in P$, for the Hilbert space ${\cal K}$ follows
from (\ref{f443q}). This completes the proof of the implication
\textbf{(i)} $\Rightarrow$ \textbf{(iii)}.

Let us note that maps $\mathfrak K_1:P\to \mathcal B(V,\mathcal H_1)$ and
$\mathfrak K_2:P\to \mathcal B(V,\mathcal H_2)$ factorize the same kernel, i.e.
\begin{equation}\label{xzsaa}
K(p,q)=\mathfrak K_1^*(p)\mathfrak K_1(q)=\mathfrak K_2^*(p)\mathfrak K_2(q)
\end{equation}
if and only if there exists a Hilbert space isomorphism
$U_{21}:\mathcal H_1\to \mathcal H_2$ such that $\mathfrak
K_2(p)=U_{21}\mathfrak K_1(p)$ for any $p\in P$. Let us define
$U_{21}$ by
\begin{equation}
U_{21}\sum_{i=1}^I \mathfrak K_1(p_i)v_i:=\sum_{i=1}^I \mathfrak K_2(p_i)v_i,
\end{equation}
where $v_i\in V$ and $p_i\in P$.

We also note that if the property \eqref{xzsaa} is fulfilled then the
Hilbert spaces $\mathcal K_1$ and $\mathcal K_2$ defined in
\eqref{ne3e} coincide, i.e., $J_1(\mathcal H_1)=J_2(\mathcal
H_2)=\mathcal K$.

Subsequently we will be interested in the case when all objects
defined above are smooth. So, further on we will assume that $P$ is a
smooth manifold.
\begin{proposition}\label{glad3}
Let $P$ be a smooth n-dimensional manifold and $V$ a
finite-dimensional complex Hilbert space. Then the following
properties are equivalent:
\begin{itemize}
\item[(a)] the positive definite kernel $K:P\times
P\rightarrow\mathcal{B}(V)$ is a smooth map.

\item[(b)] the map $\mathfrak{K}:P\rightarrow \mathcal{B}(V,{\cal
H})$ is smooth.

\item[(c)] the Hilbert space ${\cal K}\subset V^P$ defined in
\textbf{(iii)} consists of smooth functions, i.e. ${\cal K}\subset
C^\infty(P,V)$.
\end{itemize}
\end{proposition}{\it Proof}:

(a) $\Rightarrow$ (b). Let $x=(x^1,\ldots,x^n)$ be local coordinates
of $p\in P$, $y=(y^1,\ldots,y^n)$ be local coordinates of $q\in P$ and
let $(e_1,\ldots,e_n)$ be the canonical basis in $\mathbb{R}^n$.

Firstly, we prove existence of the partial derivatives. To this end
observe that for $v\in V$ and $t_1,t_2\in\mathbb{R}$ due to
(\ref{f4r54}) one has
\begin{eqnarray}\label{aaBau3w}
\lefteqn{
 \left\|\left[\frac{1}{t_1}(\mathfrak{K}(x+t_1e_i)-\mathfrak{K}(x))
-\frac{1}{t_2}(\mathfrak{K}(x+t_2e_i)-\mathfrak{K}(x))\right]v\right\|^2=
}\\
&& = \frac{1}{t_1^2}\left\langle v,[K(x+t_1e_i,x+t_1e_i)-
K(x+t_1e_i,x)-K(x,x+t_1e_i)+K(x,x)] v\right\rangle+\nonumber
\\
&&+ \frac{1}{t_2^2}\left\langle v,[K(x+t_2e_i,x+t_2e_i)-
K(x+t_2e_i,x)-K(x,x+t_2e_i)+K(x,x)] v\right\rangle+\nonumber
\\
&&+ \frac{1}{t_1t_2}\langle v,[-K(x+t_1e_i,x+t_2e_i)+
K(x+t_1e_i,x)+K(x,x+t_2e_i)-K(x,x)+\nonumber
\\
&&\ \ \ \ \ \ \ \ \ \ - K(x+t_2e_i,x+t_1e_i)+
K(x+t_2e_i,x)+K(x,x+t_1e_i)-K(x,x)] v\rangle.\nonumber
\end{eqnarray}
Since kernel $K$ is a smooth map, we have
\begin{equation}\label{ta3y1}
K(x+te_i,x)=K(x,x)+t\;\frac{\partial K}{\partial y^i}(y,x)_{|y=x}
+\frac{1}{2}t^2\;\frac{\partial^2 K}{\partial (y^i)^2}(y,x)_{|y=x}
+r_1(x;t),
\end{equation}
\begin{equation}\label{ta3y2}
K(x,x+te_i)=K(x,x)+t\;\frac{\partial K}{\partial
x^i}(y,x)_{|y=x}+\frac{1}{2}t^2\;\frac{\partial^2 K}{\partial
(x^i)^2}(y,x)_{|y=x} +r_2(x;t),
\end{equation}
\begin{eqnarray}\label{ta3y3}
\lefteqn{ K(x+t_1e_i,x+t_2e_i)=K(x,x)+t_1\frac{\partial K}{\partial
y^i}(y,x)_{|y=x}+t_2\frac{\partial K}{\partial x^i}(y,x)_{|y=x}+
}\\ \nonumber\\
&&+\frac{1}{2}t_1^2\;\frac{\partial^2 K}{\partial
(y^i)^2}(y,x)_{|y=x}+\frac{1}{2}t_2^2\;\frac{\partial^2 K}{\partial
(x^i)^2}(y,x)_{|y=x} +\frac{1}{2}t_1t_2\;\frac{\partial^2
K}{\partial y^i\partial x^i}(y,x)_{|y=x}+r_{12}(x;t_1,t_2),\nonumber
\end{eqnarray}
where
\begin{equation}\label{re3sz}
\lim_{t\rightarrow0}\frac{r_1(x;t)}{t^2}=\lim_{t\rightarrow0}\frac{r_1(x;t)}{t^2}=
\lim_{t_1,t_2\rightarrow0}\frac{r_{12}(x;t_1,t_2)}{t_1^2+t_2^2}=0.
\end{equation}
Therefore from (\ref{f4r54}) we obtain
\begin{eqnarray}\label{gaBau3w}
\lefteqn{
 \left\|\left[\frac{1}{t_1}(\mathfrak{K}(x+t_1e_i)-\mathfrak{K}(x))
-\frac{1}{t_2}(\mathfrak{K}(x+t_2e_i)-\mathfrak{K}(x))\right]v\right\|^2=
}\\
&&=\left\langle
v,\left[\frac{r_{12}(x;t_1,t_1)}{t_1^2}-\frac{r_1(x;t_1)}{t_1^2}-\frac{r_2(x;t_1)}
{t_1^2}
+\frac{r_{12}(x;t_2,t_2)}{t_2^2}-\frac{r_1(x;t_2)}{t_2^2}+
\right.\right.\nonumber \\
&&\left.\left.
-\frac{r_2(x;t_2)}{t_2^2}-\frac{r_{12}(x;t_1,t_2)}{t_1t_2}+
\frac{r_1(x;t_1)}{t_1t_2}+\frac{r_2(x;t_2)}{t_1t_2}
+
\right.\right.\nonumber \\
&&\left.\left.
-\frac{r_{12}(x;t_2,t_1)}{t_1t_2}
+\frac{r_1(x;t_2)}{t_1t_2}+\frac{r_2(x;t_1)}{t_1t_2}
\right]
v\right\rangle\xrightarrow[t_1,t_2\longrightarrow0]{}0.\nonumber
\end{eqnarray}

Since Hilbert space $V$ is finite-dimensional the above proves
existence of the partial derivatives
\begin{equation}\label{pa3r}
\frac{\partial\mathfrak{K}}{\partial x^i}(p):= \lim_{t\rightarrow
0}\frac{1}{t}(\mathfrak{K}(x+te_i)-\mathfrak{K}(x)).
\end{equation}
In order to verify continuity of $\displaystyle{
\frac{\partial\mathfrak{K}}{\partial x^i}}:P\rightarrow
\mathcal{B}(V,{\cal H}) $ note that
\begin{eqnarray}\label{s21w}
\lefteqn{
 \left\|\left[\frac{\partial\mathfrak{K}}{\partial x^i}(x+\Delta x)-
 \frac{\partial\mathfrak{K}}{\partial x^i}(x)\right]v\right\|^2=
}\\ \nonumber \\
&&=\lim_{t_1,t_2\rightarrow 0}\left\langle
v,\left[\frac{1}{t_1}(\mathfrak{K}(x+\Delta
x+te_i)-\mathfrak{K}(x+\Delta x))
-\frac{1}{t_2}(\mathfrak{K}(x+te_i)-\mathfrak{K}(x))\right]v\right\rangle=
\nonumber \\ \nonumber \\
&&=\left\langle v,\left[\frac{\partial^2 K}{\partial y^i\partial
x^i}(x+\Delta x,x+\Delta x)+\frac{\partial^2 K}{\partial y^i\partial
x^i}(x,x)-\frac{\partial^2 K}{\partial
y^i\partial x^i}(x+\Delta x,x)+\right.\right.\nonumber\\ \nonumber \\
&&\ \ \ \ \ \ \ \ \ \left.\left. -\frac{\partial^2 K}{\partial
y^i\partial x^i}(x,x+\Delta x)\right]v\right\rangle
\xrightarrow[\Delta x\rightarrow0]{}0. \nonumber
\end{eqnarray}
Continuity of partial derivatives of $\mathfrak{K}:P\rightarrow
\mathcal B(V,{\cal H})$ implies existence of its Fr\'echet derivative.
Existence of higher order Fr\'echet derivatives of
$\mathfrak{K}:P\rightarrow \mathcal B(V,{\cal H})$ can be verified in a
similar way.

Implication (b) $\Rightarrow$ (a) follows from (\ref{f4r54}).

Let us now prove that (b) $\Rightarrow$ (c). Smoothness of
$\mathfrak{K}:P\rightarrow \mathcal B(V,{\cal H})$ implies smoothness of
$\mathfrak{K}^*:P\rightarrow \mathcal B({\cal H},V)$. Thus for any
$\psi\in{\cal H}$ the function $f:P\rightarrow V$ defined by
$f(p):=\mathfrak{K}^*(p)\psi$ depends smoothly on $p\in P$.

Finally we verify implication (c) $\Rightarrow$ (a). For any $q\in
P$ and $v\in V$ one has $J(\mathfrak{K}(q)v)\in{\cal K}\subset
C^\infty(P,V)$. Thus from (\ref{ne3e}) and (\ref{f4r54}) it follows
that $K(p,q)v=\mathfrak{K}^*(p)\mathfrak{K}(q)v$ is smooth with
respect to the variable $q$. Since $K(q,p)^*=K(p,q)$ we obtain the
smooth dependence of the kernel $K$ on both arguments.

\hfill $\Box$

One of the most interesting situations arises when $P$ is a complex
analytic manifold and the scalar product in Hilbert space ${\cal
K}\subset V^P$ consisting of holomorphic functions, is defined by
the integral taken with respect to some measure $\mu$ on $P$.
Therefore Hilbert space ${\cal K}$ as well as kernel $K$ depends on
the choice of $\mu$. This case and the dependence of $K$ on $\mu$ in
particular was studied in \cite{P-W}. Many other interesting facts
and statements concerning the meaning of reproducing kernel for the
quantization can be found in \cite{B-G,G,O2,H-O,O-S}.

From now on we will assume that $P$ is a principal bundle
\begin{equation}\label{12b1}
\bfig \square/>``>`/<300,300>[G`P` `M;`` \pi `] \efig
\end{equation}
over the smooth manifold $M$ with some Lie group $G$ as the
structural group. Additionally we will assume that one has a
faithful unitary representation of $G$
\begin{equation}\label{rep32s2}
    T:G\longrightarrow \mathrm{Aut}(V)
\end{equation}
in Hilbert space $V$ and we will suppose that positive definite
kernel $K:P\times P\rightarrow\mathcal{B}(V)$ has equivariance
property
\begin{equation}\label{a21b3}
K(p,qg)=K(p,q)T(g)
\end{equation}
where $p,q\in P$ and $g\in G$. This property is equivalent to each
of the following two:
\begin{equation}\label{d443e}
\mathfrak{K}(pg)=\mathfrak{K}(p)T(g)
\end{equation}
and
\begin{equation}\label{dhu43e}
f(pg)=T(g^{-1})f(p)
\end{equation}
for $f\in{\cal K}$, where the map $\mathfrak{K}:P\rightarrow
\mathcal B(V,{\cal H})$ and Hilbert space ${\cal K}$ are defined in
\textbf{(ii)} and \textbf{(iii)}, respectively.

Using the action of $G$ on $P\times V$ defined by
\begin{equation}\label{dsa7s43}
    P\times V\ni (p,v)\mapsto (pg,T(g^{-1})v)\in P\times V
\end{equation}
one obtains $T$-associated vector bundle
\begin{equation}\label{bqw1w32}
\bfig \square/>``>`/<300,300>[V`\mathbb{V}` `M;`` \widetilde{\pi} `]
\efig
\end{equation}
over $M$ with the quotient manifold $\mathbb{V}:= (P\times V)/G$ as
its total space. The equivariance properties (\ref{a21b3}),
(\ref{d443e}) and (\ref{dhu43e}) allow us to transpose the geometric
objects defined in \textbf{(i)}, \textbf{(ii)} and \textbf{(iii)} on
the vector bundle (\ref{bqw1w32}). Note that the fiber
$\mathbb{V}_m=\widetilde{\pi}^{-1}(m)$ consists of equivalence
classes $[(p,v)]\in (P\times V)/G$ for which $\pi(p)=m$.

Given $\pi(p)=m$, $\pi(q)=n$, we define by
\begin{equation}\label{hs122q1}
K_T(m,n)([(p,v)],[(q,w)]):= \langle v, K(p,q)w\rangle,
\end{equation}
the section
\begin{equation}\label{sew2sw3q}
K_T:M\times M\longrightarrow pr_1^*\overline{\mathbb{V}}^*\otimes
pr_2^*\mathbb{V}^*
\end{equation}
of the bundle $pr_1^*\overline{\mathbb{V}}^*\otimes
pr_2^*\mathbb{V}^*\rightarrow M\times M$ which is the tensor product
of the pullbacks $pr_1^*\overline{\mathbb{V}}^*\rightarrow M\times
M$ and $pr_2^*\mathbb{V}^*\rightarrow M\times M$ of (\ref{bqw1w32}),
where $pr_i:M\times M\rightarrow M$ is the projection on the $i$-th
component of $M\times M$.

Let us define the map
$\widetilde{\mathfrak{K}}:\mathbb{V}\rightarrow {\cal H}$ by
\begin{equation}\label{desw239}
\widetilde{\mathfrak{K}}([(p,v)]):=\mathfrak{K}(p)v.
\end{equation}
Note that the kernel (\ref{hs122q1}) and the map (\ref{desw239})
are interrelated by the equality
\begin{equation}
K_T(m,n)=\iota^*_m\circ \iota_n
\end{equation}
or equivalently
\begin{equation}\label{fajr32}
K_T(m,n)([(p,v)],[(q,w)])=\langle
\widetilde{\mathfrak{K}}([(p,v)])\mid
\widetilde{\mathfrak{K}}([(q,w)])\rangle,
\end{equation}
where $\iota_m:=\widetilde{\mathfrak K}_{|\mathbb V_m}:\mathbb
V_m\to \mathcal H$.

We also can realize Hilbert space ${\cal H}$ by smooth sections of
vector bundles $\widetilde{\pi}^*:\mathbb{V}^*\rightarrow M$ or
$\widetilde{\pi}:\mathbb{V}\rightarrow M$. The first realization is
given by the anti-linear monomorphism of vector spaces $I^*:{\cal
H}\rightarrow C^\infty(M,\mathbb{V}^*)$ defined for $\psi\in{\cal
H}$ in the following way
\begin{equation}\label{dz4e3}
    I^*(\psi)(\pi(p))([(p,v)]):=\langle\psi \mid\mathfrak{K}(p)v\rangle
    =\langle J(\psi)(p),v\rangle.
\end{equation}
The second realization is as follows
\begin{equation}\label{wie5l}
    I(\psi)(\pi(p)):=[(p,\mathfrak{K}(p)^*\psi) ]=[(p,J(\psi)(p)) ].
\end{equation}
The map $I:{\cal H}\rightarrow C^\infty(M,\mathbb{V})$ is also a
linear monomorphism of vector spaces. By $C^\infty(M,\mathbb{V}^*)$
and $C^\infty(M,\mathbb{V})$ we denote vector spaces of smooth
sections of the bundles $\widetilde{\pi}^*:\mathbb{V}^*\rightarrow
M$ and $\widetilde{\pi}:\mathbb{V}\rightarrow M$, respectively.

The restriction $K_{T|\Delta}$ of the kernel $K_T$ to the diagonal
$\Delta:=\{(m,n)\in M\times M: m=n\}$ determines a semi-positive
definite Hermitian structure $H_K:=K_{T|\Delta}$ on the bundle
$\widetilde{\pi}:\mathbb{V}\rightarrow M$. For subsequent
considerations we will need to deal with the positive definite
Hermitian structure. So, further on we will assume that linear
operator $K(p,p)$ is invertible for every $p\in P$. The last
condition is equivalent to the condition
$\ker\mathfrak{K}(p)=\{0\}$, for $p\in P$. Recall that we assumed
$\dim V=: N<\infty$.

Apart from the Hermitian structure $H_K$ the positive Hermitian kernel $K$
defines on $P$ a ${\cal B}(V)$-valued differential one-form
\begin{equation}\label{je3sze}
    \vartheta(p):=
    (\mathfrak{K}(p)^*\mathfrak{K}(p))^{-1}\mathfrak{K}(p)^*d\mathfrak{K}(p)
    =K(p,p)^{-1}d_qK(p,q)_{|q=p},
\end{equation}
where $d_q$ denotes the exterior derivative with respect to the
variable $q$.

Let $g_\mathfrak{X}(t):=\exp(t\mathfrak{X})$ denote a one-parameter
group generated by an element $\mathfrak{X}\in \mathfrak{g}=T_eG$ of
Lie algebra of G. By $\xi_\mathfrak{X}\in C^\infty(P,TP)$ we denote
vector field tangent to the vertical flow $p\mapsto
pg_\mathfrak{X}(t)$. From (\ref{je3sze}) and (\ref{d443e}) it
follows that
\begin{equation}\label{k87on5}
    (\xi_\mathfrak{X}\llcorner\vartheta)(p)=
    (\mathfrak{K}(p)^*\mathfrak{K}(p))^{-1}
    \mathfrak{K}(p)^*(\xi_\mathfrak{X}\llcorner d\mathfrak{K})(p)
    =\frac{d}{dt}T(g_\mathfrak{X}(t))_{|t=0}=DT(e)(\mathfrak{X})
\end{equation}
and
\begin{equation}\label{a32dj5}
    \vartheta(pg)=T(g^{-1})\vartheta(p)T(g),
\end{equation}
where $DT(e):\mathfrak{g}\rightarrow {\cal B}(V)$ is the derivative
of the representation (\ref{rep32s2}) taken at the unit element
$e\in G$.

Taking (\ref{k87on5}) and (\ref{a32dj5}) into account we conclude
from
\begin{equation}\label{pr32e1}
    \langle
    v,K(p,p)\vartheta(p)w\rangle+\langle\vartheta(p)v,K(p,p)w\rangle=
    d\langle v,K(p,p)w\rangle
\end{equation}
that $\vartheta\in C^\infty(P,T^*P\otimes{\cal B}(V))$ is the one-form
of a metric connection $\nabla_K$ consistent with the Hermitian
structure $H_K$.

Now let us consider the Grassmannian $Gr(N,{\cal H})$ of
$N$-dimensional Hilbert subspaces of ${\cal H}$ and the tautological
vector bundle
\begin{equation}\label{n1f554}
\bfig \square/>``>`/<300,300>[V`\mathbb{E}` `Gr(N,{\cal H});``
pr_1 `] \efig
\end{equation}
where total space of (\ref{n1f554}) is defined by
\begin{equation}\label{dew332w}
    \mathbb{E}:=\left\{(z,\psi)\in Gr(N,{\cal
H})\times \mathcal H:\psi\in z \right\}.
\end{equation}
The scalar product in ${\cal H}$ defines an Hermitian structure
$H_{\mathbb{E}}$ on the vector bundle $pr_2:\mathbb{E}\rightarrow
Gr(N,{\cal H})$ in the canonical way. There is also a unique
connection on this bundle
\begin{equation}\label{conf4s}
    \nabla_{\mathbb{E}}:C^\infty(Gr(N,{\cal
    H}),\mathbb{E})\rightarrow C^\infty(Gr(N,{\cal H}),\mathbb{E}\otimes
    T^*(Gr(N,{\cal H}))),
\end{equation}
consistent with $H_{\mathbb{E}}$ and complex analytic structures of
$Gr(N,{\cal H})$. See \cite{K-N}, Volume 1, Chapter 2, for the
definition of such connections.

Since $\ker\widetilde{\mathfrak{K}}_{|\mathbb{V}_m}=\{0\}$ one has a
map ${\cal K}:M\rightarrow Gr(N,{\cal H})$ defined by
\begin{equation}\label{h4ornot}
    {\cal K}(m):=\widetilde{\mathfrak{K}}(\mathbb{V}_m).
\end{equation}
Thus one has also the following vector bundle morphism
\begin{equation}\label{nde6not}
\bfig \square/>`>`>`>/<700,500>[
\mathbb{V}`\mathbb{E}`M`Gr(N,{\cal
H});\widetilde{\widetilde{\mathfrak{K}}}`\widetilde{\pi}`pr_1`{\cal
K}] \efig
\end{equation}
where
\begin{equation}\label{h5ornot}
    \widetilde{\widetilde{\mathfrak{K}}}([(p,v)]):=
    (\widetilde{\mathfrak{K}}([(p,v)]),{\cal K}(m)
    ).
\end{equation}
Taking into account the definitions of $H_K$ and $\nabla_K$ we find
that they are pullbacks $H_K={\cal K}^*H_{\mathbb{E}}$ and
$\nabla_K={\cal K}^*\nabla_{\mathbb{E}}$ of $H_{\mathbb E}$ and
$\nabla_{\mathbb{E}}$, respectively. So, (\ref{nde6not}) gives a
vector bundle morphism which preserves Hermitian and connection
structures.

It is important to mention that in geometric models of physical
systems one considers a bundle $\tilde \pi:\mathbb V\to M$ as the
space of states, where the fibers $\tilde\pi^{-1}(m)$, $m\in M$
describe internal degrees of freedom and the base manifold $M$ is
responsible for the external degrees of freedom of the system.
Within such an interpretation a positive definite kernel $K(p,q)$
after normalization can be considered as a transition amplitude
\begin{equation}\label{amp}
a([(p,v)],[(q,w)]):=\frac{\langle v, K(p,q) w \rangle}
{\langle v, K(p,p) v\rangle^{\frac12}\langle w, K(q,q) w\rangle^{\frac12}}
\end{equation}
between the states
$\widetilde{\mathfrak{K}}([(p,v)]),\widetilde{\mathfrak{K}}([(q,w)])\in
{\cal H}$. Let us mention here that from physical point of view
transition amplitude \eqref{amp} is the most fundamental object
which is usually obtained in an experimental way (see \cite{F-L-S}
Chapter 3).

Finally let us make some comments:

i) The maps $\mathfrak K:P\to \mathcal B(V,\mathcal H)$,
$\tilde{\mathfrak K}:\mathbb V\to\mathcal H$ and kernel $K_T$ define
 each other explicitly.

ii) The scalar product $\langle\cdot|\cdot\rangle$ of $\mathcal H$
defines positive definite kernel $K_{\mathbb E}:Gr(N,\mathcal
H)\times Gr(N,\mathcal H)\to pr^*_1\bar{\mathbb E}^*\otimes
pr^*_2\mathbb E^*$ on the tautological bundle $\pi:\mathbb E\to
Gr(N,\mathcal H)$ by
\begin{equation}
K_E((z_1,\psi_1),(z_2,\psi_2)):=\langle \psi_1|\psi_2\rangle.
\end{equation}

iii) The pullback of the kernel $K_{\mathbb E}$ on the vector bundle
$\tilde \pi:\mathbb V\to M$ by the map $\tilde{\mathfrak K}:\mathbb
V\to \mathcal H$ gives the kernel $K_T$ defined in \eqref{hs122q1}.
Therefore one can consider $(\mathbb E,K_{\mathbb E})$ as the
universal object in the category of vector bundles $(\mathbb V,K_T)$
with the fixed positive definite kernel $K_T$.

Similarly, the prequantum bundle $(\mathbb E,\nabla_{\mathbb E},
H_{\mathbb E})$, where $\nabla_{\mathbb E}$ and $H_{\mathbb E}$ are
the  connection and the Hermitian structure on $\pi:\mathbb E\to
Gr(N,\mathcal H)$ defined by kernel $K_{\mathbb E}$, is the
universal object in the category of prequantum bundles $(\mathbb
V,\nabla_K,H_K)$ defined by $(\mathbb V,K_T)$. The relationship
between $(\mathbb V,K_T)$ and $(\mathbb V,\nabla_K, H_H)$ has
functorial character.

In subsequent sections we will call $\tilde{\mathfrak K}:\mathbb V\to
\mathcal H$ as well as $\mathfrak K:P\to \mathcal B(V,\mathcal H)$
the \emph{coherent state map}. See \cite{O1} and \cite{O2} for a
physical motivation of this terminology.

\section{One-parameter groups of automorphisms and prequantization}

In this section we introduce various Lie algebras, see Proposition
3.1, Proposition 3.2 and Proposition 3.3, the elements of which
generate the one--parameter group of automorphisms of the principal
bundle $\pi: P\to M$. We will show the special importance of the Lie
algebras appearing in the short exact sequence (\ref{dok81cia}).
This is important because this sequence generalizes the exact
sequence of Lie algebras
\begin{equation}
0\rightarrow \mathbb{R}\rightarrow C^{\infty}(M, \mathbb{R})
\rightarrow H_0(M,\omega)\rightarrow 0, \nonumber
\end{equation}
for the symplectic manifold $M$ with symplectic form $\omega$,
where
$H_0(M, \omega)$ is the Lie algebra of Hamiltonian vector fields. We
also introduce here equation (\ref{ham87i}) which is a natural
generalization of Hamiltonian equation for the case of general gauge
group $G$.

Now we extend Kostant-Souriau prequantization procedure
on the case of general principal G-bundle $\pi:P\rightarrow M$ with
a fixed $DT(e)(\mathfrak{g})$-valued connection form $\vartheta\in
C^\infty_G(T^*P\otimes DT(e)(\mathfrak{g}))$.

Let $\xi\in C^\infty(P,TP)$ be the vector field tangent to the flow
of automorphisms $\tau:(\mathbb{R},+)\rightarrow
\mbox{Aut}(P,\vartheta)$ of the principal bundle (\ref{12b1})
\begin{equation}\label{juh6qq}
    \tau_t(pg)=\tau_t(p)g,
\end{equation}
where $g\in G$ and $p\in P$, which preserve the connection form
$\vartheta$
\begin{equation}\label{nie48kon}
    \tau_t^*\vartheta=\vartheta.
\end{equation}
Then one has
\begin{equation}\label{pr4es}
    \xi(pg)=DR_g(p)\xi(p),
\end{equation}
and
\begin{equation}\label{lie5dre}
{\cal L}_\xi\vartheta=0,
\end{equation}
where $R_g(p):=pg$, while $DR_g(p)$ is the derivative of $R_g$ at
$p$ and ${\cal L}_\xi$ is Lie derivative with respect to $\xi$.

The space of vector fields satisfying (\ref{pr4es}) will be denoted
by $C^\infty_G(P,TP)$. By ${\cal E}_G^0\subset C_G^\infty(P,TP)$ we
denote the subspace consisting of those $\xi\in C_G^\infty(P,TP)$
which satisfy (\ref{lie5dre}).

Recall that the covariant differential $\textbf{D}\varphi$ of
$DT(e)(\mathfrak{g})$-valued pseudotensorial r-form $\varphi$ on $P$
is defined as $\textbf{D}\varphi(\xi_1,\ldots,\xi_{r+1}):=d\varphi(
pr_{hor}\xi_1,\ldots,pr_{hor}\xi_{r+1})$, where
$pr_{hor}(p):T_pP\rightarrow T_p^hP$ is projection associated with
the decomposition $T_pP= T_p^hP\oplus T_p^vP$ of the tangent space
$T_pP$ in the horizontal and vertical parts taken with respect to
the connection form (\ref{je3sze}). In particular, for connection
1-form $\vartheta$ and $DT(e)(\mathfrak{g})$-valued pseudotensorial
0-form, i.e. $DT(e)(\mathfrak{g})$-valued function such that
\begin{equation}\label{ps3eu}
    F(pg)= T(g^{-1})F(p)T(g),
\end{equation}
one has
\begin{equation}\label{str4e}
    \textbf{D}\vartheta=d\vartheta+\frac{1}{2} [\vartheta,\vartheta],
\end{equation}
\begin{equation}\label{kodyfu}
    \textbf{D}F=dF+[\vartheta,F],
\end{equation}
where (\ref{str4e}) is the structure equation for the curvature form
$\Omega:=\textbf{D}\vartheta$, (see for example \cite{K-N}). In the
subsequent we will use the notation taken from \cite{K-N}, Volume 1
Chapter 2.

By $C^\infty_G(P,DT(e)(\mathfrak{g}))$ we denote the space of
$DT(e)(\mathfrak{g})$-valued functions satisfying condition
(\ref{ps3eu}).

Now let us consider the vector space ${\cal P}_G$ which by
definition consists of pairs $(F,\xi)\in
C^\infty_G(P,DT(e)(\mathfrak{g}))\times C^\infty_G(P,TP)$ such that
\begin{equation}\label{ham87i}
    \xi\llcorner\Omega=\textbf{D}F.
\end{equation}

\begin{proposition}\label{nowe1}
For $(F,\xi),(G,\eta)\in{\cal P}_G$ we have
\begin{equation}\label{dl76ra}
{\cal
L}_{[\xi,\eta]}\vartheta=\textbf{D}(\{F,G\}+\vartheta([\xi,\eta])),
\end{equation}
where
\begin{equation}\label{na3pl}
    \{F,G\}:=2\Omega(\xi,\eta)+\textbf{D}G(\xi)-\textbf{D}F(\eta)+[F,G].
\end{equation}
\end{proposition}{\it Proof}:

Due to the identity
\begin{equation}\label{lie32der}
{\cal L}_\xi\vartheta=\xi\llcorner\Omega+\textbf{D}(\vartheta(\xi)),
\end{equation}
the condition (\ref{ham87i}) is equivalent to
\begin{equation}\label{po8li}
    {\cal L}_\xi\vartheta=\textbf{D}(F+\vartheta(\xi)).
\end{equation}

Using (\ref{str4e}), (\ref{kodyfu}), (\ref{po8li}) and definition
(\ref{na3pl}) we obtain
\begin{equation}
{\cal L}_{[\xi,\eta]}\vartheta={\cal L}_{\xi}{\cal
L}_{\eta}\vartheta-{\cal L}_{\eta}{\cal L}_{\xi}\vartheta= {\cal
L}_{\xi}(\textbf{D}(G+\vartheta(\eta)))-{\cal
L}_{\eta}(\textbf{D}(F+\vartheta(\xi)))=\nonumber
\end{equation}
\begin{equation}
={\cal
L}_{\xi}\left(d(G+\vartheta(\eta))+[\vartheta,G+\vartheta(\eta)]\right)
-{\cal
L}_{\eta}(d(F+\vartheta(\xi))+[\vartheta,F+\vartheta(\xi)])=\nonumber
\end{equation}
\begin{equation}
=d(\xi(G)+\xi(\vartheta(\eta))-\eta(F)-\eta(\vartheta(\xi))+ {\cal
L}_{\xi}\left([\vartheta,G+\vartheta(\eta)]\right) -{\cal
L}_{\eta}([\vartheta,F+\vartheta(\xi)])=\nonumber
\end{equation}
\begin{equation}
=d(\xi(G)+\xi(\vartheta(\eta))-\eta(F)-\eta(\vartheta(\xi))+\nonumber
\end{equation}
\begin{equation}
+[\textbf{D}(F+\vartheta(\xi)),G+\vartheta(\eta)]+
[\vartheta,\xi(G)+\xi(\vartheta(\eta))] -\nonumber
\end{equation}
\begin{equation}
-[\textbf{D}(G+\vartheta(\eta)),F+\vartheta(\xi)]-
[\vartheta,\eta(F)+\eta(\vartheta(\xi))] =\nonumber
\end{equation}
\begin{equation}
=\textbf{D}(\xi(G)+\xi(\vartheta(\eta))-\eta(F)-\eta(\vartheta(\xi))+\nonumber
\end{equation}
\begin{equation}
+[\textbf{D}(F+\vartheta(\xi)),G+\vartheta(\eta)]
-[\textbf{D}(G+\vartheta(\eta)),F+\vartheta(\xi)] =\nonumber
\end{equation}
\begin{equation}
=\textbf{D}(\xi(G)+\xi(\vartheta(\eta))-\eta(F)-\eta(\vartheta(\xi))+
[F+\vartheta(\xi),G+\vartheta(\eta)]) =\nonumber
\end{equation}
\begin{equation}
=\textbf{D}(\xi(G)-\eta(F)+[F+\vartheta(\xi),G+\vartheta(\eta)]+2\Omega(\xi,\eta)
-[\vartheta(\xi),\vartheta(\eta)]+\vartheta([\xi,\eta]))=\nonumber
\end{equation}
\begin{equation}
=\textbf{D}((\textbf{D}G)(\xi)-(\textbf{D}F)(\eta)+[F,G]+2\Omega(\xi,\eta)
+\vartheta([\xi,\eta]))=\nonumber
\end{equation}
\begin{equation}
=\textbf{D}(\{F,G\}+\vartheta([\xi,\eta]) ).\nonumber
\end{equation}

\hfill $\Box$

Let us note that bracket (\ref{na3pl}) could be defined in the
following equivalent ways
\begin{equation}\label{na3pl22222}
 \{F,G\}:=-2\Omega(\xi,\eta)+[F,G]=\textbf{D}G(\xi)+[F,G]=
 -\textbf{D}F(\eta)+[G,F].
 \end{equation}

\begin{proposition}\label{nowe12}
The space ${\cal P}_G$ with the bracket
\begin{equation}\label{bra32k}
[\![(F,\xi),(G,\eta)]\!]:=(\{F,G\},[\xi,\eta])
\end{equation}
is a Lie algebra.
\end{proposition}{\it Proof}:

For arbitrary $(F,\xi),(G,\eta), (H,\lambda)\in{\cal P}_G$ by direct
calculations we obtain
\begin{eqnarray}\label{Arqq2sd}
 &\{F,\{G,H\}\}= \textbf{D}((\textbf{D}H)(\eta)-(\textbf{D}G)(\lambda)+[G,H]+
    2\Omega(\eta,\lambda))(\xi)+&\\
 &+[F,(\textbf{D}H)(\eta)-(\textbf{D}G)(\lambda) +[G,H]+
 2\Omega(\eta,\lambda)].&\nonumber
 \end{eqnarray}
Adding the cyclic permutations of both sides of (\ref{Arqq2sd}) we
find that
\begin{equation}\label{jac5oq}
    \{F,\{G,H\}\}+\{G,\{H,F\}\}+\{H,\{F,G\}\}=0.
\end{equation}
Thus we conclude that bracket (\ref{bra32k}) satisfies Jacobi
identity and hence $({\cal P}_G,[\![\cdot,\cdot]\!] )$ is a Lie
algebra.

\hfill $\Box$

Let ${\cal E}_G$ be the Lie algebra of vector fields $\xi\in
C^\infty_G(P,TP)$ for which there exist $F\in
C^\infty_G(P,DT(e)(\mathfrak{g}))$ such that $(F,\xi)\in{\cal P}_G$.
Denote by ${\cal N}_G$ the set of $F\in
C^\infty_G(P,DT(e)(\mathfrak{g}))$ for which $\textbf{D}F=0$. One
has the following exact sequence of Lie algebras
\begin{equation}\label{ci8do}
    0\rightarrow{\cal N}_G\xrightarrow[]{\iota_1}{\cal P}_G
    \xrightarrow[]{pr_2}{\cal
    E}_G\rightarrow0,
\end{equation}
where
\begin{equation}\label{pij7y}
\iota_1(F):=(F,0),\;\;\;pr_2(F,\xi):=\xi.
\end{equation}

From now on we will assume that $M$ is a connected manifold and
denote by $P(p)$ the set of elements of $P$ which one can join with
$p$ by curves which are horizontal with respect to the connection $\vartheta$.
By $G(p)$ we denote the subgroup $G(p)\subset G$ consisting of those
$g\in G$ for which $pg\in P(p)$, i.e. $G(p)$ is the holonomy group
based at $p$. Let us recall (e.g. see \cite{K-N}) that for connected
base manifold $M$ all holonomy groups $G(p)$ and their Lie algebras
$\mathfrak{g}(p)$ are conjugate in $G$ and $\mathfrak{g}$,
respectively. Recall also that Lie algebra $DT(e)(\mathfrak{g}(p))$
is generated by $\Omega_{p'}(X(p'),Y(p'))$, where $p'\in P(p)$ and
$X(p'),Y(p')\in T_{p'}P$.

Taking this into account we conclude from condition (\ref{ham87i})
that for $(F,\xi)\in{\cal P}_G$ function $F$ takes values $F(p')$ in
$\mathfrak{g}(p)$ if $p'\in P(p)$. In the special case when $F\in{\cal
N}_G$, i.e. when $\textbf{D}F=0$, function $F$ is constant on $P(p)$
and $F(p)\in DT(e)(\mathfrak{g}(p))\cap DT(e)(\mathfrak{g}'(p))$,
where $\mathfrak{g}'(p)$ is the centralizer of Lie subalgebra
$\mathfrak{g}(p)$ in $\mathfrak{g}$.

For the sake of completeness let us note that
\begin{equation}\label{dr4er}
{\cal
L}_\xi\Omega=[\Omega,F+\vartheta(\xi)]=\textbf{D}^2(F+\vartheta(\xi))=
\textbf{D}{\cal L}_\xi\vartheta,
\end{equation}
for $(F,\xi)\in{\cal P}_G$.

It follows from (\ref{lie32der}) that ${\cal E}_G^0\subset{\cal
E}_G$. Thus we can consider the subspace ${\cal P}_G^0\subset {\cal
P}_G$ of such elements $(F,\xi)\in {\cal P}_G$ that $\xi\in{\cal
E}_G^0$ and $F=F_0-\vartheta(\xi)$, where $\textbf{D}F_0=0$. For
$\xi,\eta\in{\cal E}_G^0$ we have
\begin{equation}\label{fspr777}
\vartheta([\xi,\eta])=2\Omega(\xi,\eta)-[\vartheta(\xi),\vartheta(\eta)].
\end{equation}
Thus for $(F,\xi),(G,\eta)\in{\cal P}_G^0$ we obtain
\begin{eqnarray}
 &[\![(F,\xi),(G,\eta)]\!]=
[\![(F_0,0)+(-\vartheta(\xi),\xi),(G_0,0)+(-\vartheta(\eta),\eta)]\!]=
&\nonumber\\
 &=[\![(-\vartheta(\xi),\xi),(-\vartheta(\eta),\eta)]\!]=
 (\{\vartheta(\xi),\vartheta(\eta) \}, [\xi,\eta])=
 (-\vartheta([\xi,\eta]),[\xi,\eta]).&\label{Arqq2}
 \end{eqnarray}

From (\ref{dl76ra}) and (\ref{Arqq2}) we see that ${\cal P}_G^0$ is
a Lie subalgebra of ${\cal P}_G$ which contains the ideal ${\cal
N}_G$.

Summing up we accumulate the above facts in the following diagram
\begin{equation}\label{Baz}
\begin{array}{ccccccccc}
0&\rightarrow&{\cal N}_G&\xrightarrow[]{\iota_1}&{\cal P}_G&
    \xrightarrow[]{pr_2}&{\cal
    E}_G&\rightarrow&0,\\
& &\uparrow& & \uparrow& & \uparrow& & \\
0&\rightarrow&{\cal N}_G&\xrightarrow[]{\iota_1}&{\cal P}_G^0&
    \xrightarrow[]{pr_2}&{\cal
    E}_G^0&\rightarrow&0,
\end{array}
\end{equation}
where horizontal arrows form the exact sequences of Lie algebras and
vertical arrows are Lie algebra monomorphisms.

In order to describe Lie algebra ${\cal P}_G^0$ we define a linear
monomorphism $\Phi:{\cal E}_G^0\rightarrow{\cal P}_G^0$ by
\begin{equation}\label{fi76j}
    \Phi(\xi):=(-\vartheta(\xi),\xi).
\end{equation}
It follows from (\ref{Arqq2}) that $\Phi$ is a monomorphism of Lie
algebras. Due to (\ref{po8li}) one has the decomposition
\begin{equation}\label{su7pr}
{\cal P}_G^0=\iota_1({\cal N}_G)\oplus\Phi({\cal E}_G^0)
\end{equation}
of ${\cal P}_G^0$ into the direct sum of Lie subalgebra $\Phi({\cal
E}_G^0)$ and ideal $\iota_1({\cal N}_G)$ of central elements of
${\cal P}_G^0$. Hence we conclude that the lower exact sequence of
Lie algebras in (\ref{Baz}) is trivial.

Now let us define the following Lie subalgebra
\begin{equation}\label{ha3m}
{\cal H}_G^0:= D\pi ({\cal E}_G^0),
\end{equation}
of $C^\infty(M,TM)$, where $D\pi:TP\rightarrow TM$ is the tangent
map of the bundle map $\pi:P\rightarrow M$. We also define
 ${\cal F}_G^0\subset
C^\infty_G(P,DT(e)(\mathfrak{g}))\times{\cal H}_G^0$ as the vector subspace
consisting of
elements $(F,X)\in C^\infty_G(P,DT(e)(\mathfrak{g}))\times{\cal
H}_G^0$ which satisfy the condition
\begin{equation}\label{h71am87i}
    X^*\llcorner\Omega=\textbf{D}F,
\end{equation}
where $X^*$ is the horizontal lift of $X$ with respect to
$\vartheta$. From condition (\ref{h71am87i}) and identity
(\ref{lie32der}) it follows that
\begin{equation}\label{roz312w}
    \xi:=X^*-F^*\in{\cal E}_G^0 ,
\end{equation}
where $F^*$ is a vertical field defined by the function $F\in
C^\infty_G(P,DT(e)(\mathfrak{g}))$ in the following way
\begin{equation}\label{gw32w}
    (F^*f)(p)= \frac{d}{dt}f(p\exp(tF'(p)))_{|t=0},
\end{equation}
where $f\in C^\infty(P)$ and function $F':P\rightarrow\mathfrak{g}$
is such that $DT(e)(F'(p))=F(p)$. Note that (\ref{roz312w}) gives
the decomposition of $\xi=\xi^h+\xi^v$ on the horizontal $\xi^h=X^*$
and vertical $\xi^v=-F^*$ components.

On the other side decomposing $\xi=\xi^h+\xi^v\in{\cal E}_G^0$ on
the horizontal and vertical parts we define
\begin{equation}\label{re41pop}
    (F,X)=(-\vartheta(\xi^v),D\pi(\xi^h))\in{\cal F}_G^0.
\end{equation}
Summing up the above facts we formulate the following
\begin{proposition}\label{prop:3.1}
The relations (\ref{roz312w}) and (\ref{re41pop}) define a Lie
algebras isomorphism between $({\cal E}_G^0, [\cdot,\cdot])$ and
$({\cal F}_G^0,\{\!\!\{\cdot,\cdot\}\!\!\})$, where the Lie bracket
of $(F,X),(G,Y)\in{\cal F}_G^0$ is defined by
\begin{equation}\label{lie423spr}
\{\!\!\{(F,X),(G,Y)\}\!\!\}:=(-2\Omega(X^*,Y^*)+[F,G],[X,Y]).
\end{equation}

One has the following exact sequence of Lie algebras
\begin{equation}\label{dok81cia}
    0\rightarrow{\cal N}_G\xrightarrow[]{\iota_1}{\cal F}_G^0
    \xrightarrow[]{pr_2}{\cal
    H}_G^0\rightarrow0,
\end{equation}
where $\iota_1(F):=(F,0)$ and $pr_2(F,X):=X$.
\end{proposition}

The integration of the horizontal part $\xi^h=X^*$ of $\xi\in{\cal
E}_G^0$ gives the flow $\{\tau_t^h\}_{t\in\mathbb R}$ being the
horizontal lift of the flow
\begin{equation}\label{na3ba}
    \sigma:(\mathbb{R},+)\longrightarrow \mbox{Diff}(M)
\end{equation}
defined by the projection of $\{\tau_t\}_{t\in\mathbb R}$ on the
base $M$ of the principal bundle $P$. The vector field $X\in{\cal
H}_G^0$ is the velocity vector field of $\{\sigma_t\}_{t\in\mathbb
R}$.

Since $\{\tau_t\}_{t\in\mathbb R}$ and $\{\tau_t^h\}_{t\in\mathbb
R}$ are the liftings of $\{\sigma_t\}_{t\in\mathbb R}$ there exists
 a $G$-valued cocycle on $P$, namely a map $c:\mathbb R\times
P\rightarrow G$ such that
\begin{equation}\label{co5cy1}
    c(t+s,p)=c(t,\tau^h_s(p))c(s,p)=c(s,p)c(t,\tau_s(p)),
\end{equation}
\begin{equation}\label{co5cy2}
    c(t,pg)=g^{-1}c(t,p)g,
\end{equation}
which intertwines both flows
\begin{equation}\label{int6r}
    \tau_t(p)=\tau_t^h(p)c(t,p).
\end{equation}

Applying the representation $T:G\rightarrow \mbox{Aut}(V)$
(see(\ref{rep32s2})) to (\ref{co5cy1}) and subsequently
differentiating (\ref{co5cy1}) with respect to the parameter $t$ at
$t=0$ we obtain differential equation
\begin{equation}\label{na43er}
    \frac{d}{ds}T(c(s,p))=T(c(s,p))\frac{d}{dt}T(c(t,\tau_s(p))_{|t=0}
\end{equation}
with initial condition $T(c(0,p))=1\!\!1$.

In order to solve (\ref{na43er}) note that from definition
(\ref{je3sze}) one has
\begin{equation}
\vartheta_p(\xi(p))=K(p,p)^{-1}\lim_{\Delta
t\rightarrow0}\frac{1}{\Delta t}
 [K(p,\tau_{\Delta t}(p)-K(p,p)]=
\end{equation}
\begin{equation}
= K(p,p)^{-1}   \lim_{\Delta t\rightarrow0}\frac{1}{\Delta t}
 [K(p,\tau_{\Delta t}^h(p)c(\Delta t,p))-K(p,p)]=
\nonumber
\end{equation}
\begin{equation}
= K(p,p)^{-1}   \lim_{\Delta t\rightarrow0}\frac{1}{\Delta t}
 [K(p,\tau_{\Delta t}^h(p)c(\Delta t,p))-K(p,pc(\Delta t,p))+
 K(p,pc(\Delta t,p))-K(p,p)]=\nonumber
\end{equation}
\begin{equation}
= K(p,p)^{-1}   \lim_{\Delta t\rightarrow0}\frac{1}{\Delta t}
 [K(p,\tau_{\Delta t}^h(p)c(\Delta t,p))-K(p,pc(\Delta t,p))]+\nonumber
\end{equation}
\begin{equation}
+K(p,p)^{-1}   \lim_{\Delta t\rightarrow0}\frac{1}{\Delta t}
 [K(p,pc(\Delta t,p))-K(p,p)]=\nonumber
\end{equation}
\begin{equation}
=\vartheta_p(\xi^h(p))+K(p,p)^{-1} K(p,p)
 \lim_{\Delta t\rightarrow0}\frac{1}{\Delta t}
 [T(c(\Delta t,p))-1\!\!1]=\frac{d}{dt}T(c(t,p))_{|t=0}.\nonumber
\end{equation}
By virtue of
\begin{equation}\label{sp24ra}
{\cal L}_\xi(\xi\llcorner\vartheta)=
[\xi,\xi]\llcorner\vartheta+\xi\llcorner({\cal L}_\xi\vartheta)=0,
\end{equation}
 we obtain
\begin{equation}\label{oc6zy}
\frac{d}{dt}T(c(t,\tau_s(p))_{|t=0}=\vartheta(\xi)(\tau_s(p))=
\vartheta(\xi)(p).
\end{equation}

Now, solving equation
\begin{equation}\label{na43ewq}
    \frac{d}{ds}T(c(s,p))=T(c(s,p))\vartheta(\xi)(p)
\end{equation}
with $T(c(0,p))=1\!\!1$ we get
\begin{equation}\label{ro7zw}
T(c(t,p))=\exp(t\vartheta(\xi)(p))=\exp(-tF(p)).
\end{equation}
Taking (\ref{ro7zw}) into account and making use of the fact that
$\{\sigma_t\}_{t\in\mathbb R}$ is defined by $X\in {\cal H}_G^0$ we
conclude from (\ref{int6r}) that $\{\tau_t\}_{t\in\mathbb R}$ is
determined in a unique way by $(F,X)\in{\cal F}_G^0$.

In the case when $G=U(1)$, $\dim V=1$ and the curvature $\Omega$ is
a non-singular 2-form equation (\ref{h71am87i}) reduces to the
Hamilton equation with $F\in C_G^\infty(P)\cong C^\infty(M)$ as a
Hamiltonian (total energy function). So, it is natural to consider
(\ref{h71am87i}) as a generalization of the Hamilton equations to
the case of general gauge group $G$. If $\Omega$ is non-singular one
can consider the vector field $X_F\in {\cal H}_G^0$ defined by $F\in
pr_1({\cal F}_G^0)$
as the Hamiltonian field. By definition the space $pr_1({\cal
F}_G^0)$ consists  of $F\in C^\infty_G(P,DT(e)(\mathfrak{g}))$ such
that $(F,X)\in{\cal F}_G^0$. Note that $pr_1({\cal
F}_G^0)\varsubsetneq C^\infty_G(P,DT(e)(\mathfrak{g}))$ in general
case. However, if $G=U(1)$ one has equality $pr_1({\cal
F}_G^0)=C_G^\infty(P)\cong C^\infty(M)$.

Now we generalize the Kostant--Souriau prequantization procedure to
the general gauge group case. To this end let us consider the space
$C^\infty_G(P,V)$ of $V$-valued functions $f:P\rightarrow V$
equivariant with respect to the gauge group $G$
\begin{equation}\label{eu76i}
    f(pg)=T(g^{-1})f(p),
\end{equation}
where $p\in P$ and $g\in G$.

The flow $\{\tau_t \}_{t\in\mathbb
R}\subset \mbox{Aut}(P,\vartheta)$ defines a one-parameter group
$\Sigma_t:C^\infty_G(P,V)\rightarrow C^\infty_G(P,V)$ of
automorphisms of the vector space $C^\infty_G(P,V)$
\begin{equation}\label{fl87w3}
    (\Sigma_tf)(p):=f(\tau_{-t}(p)).
\end{equation}

Defining the flow
\begin{equation}\label{wi8we}
    \widetilde{\tau}_t[(p,v)]:=[(\tau_t(p),v)]
\end{equation}
on the vector bundle $\widetilde{\pi}:\mathbb{V}\rightarrow M$ one
obtains the one-parameter group
$\widetilde{\Sigma}_t:C^\infty(M,\mathbb{V})\rightarrow
C^\infty(M,\mathbb{V})$ acting on the sections $\psi\in
C^\infty(M,\mathbb{V})$ in the following way
\begin{equation}\label{je43je}
(\widetilde{\Sigma}_t\psi)(\pi(p)):=\widetilde{\tau}_t
\psi(\sigma_{-t}\circ\pi(p))=\widetilde{\tau}_t
\psi(\pi(\tau_{-t}(p)))= \widetilde{\tau}_t
\psi(\pi(\tau^h_{-t}(p))).
\end{equation}

The isomorphism ${\cal R}:C^\infty_G(P,V)\xrightarrow[]{\sim}
C^\infty(M,\mathbb{V})$
defined by
\begin{equation}\label{er2re}
    ({\cal R}f)(\pi(p)):=[(p,f(p))]
\end{equation}
intertwines the flows (\ref{fl87w3}) and (\ref{je43je})
\begin{equation}\label{sp98la}
{\cal R}\circ \Sigma_t=\widetilde{\Sigma}_t  \circ{\cal R}
\end{equation}
and moreover
\begin{equation}
\mathcal R\circ J=I,
\end{equation}
where $I:\mathcal H\to C^\infty(M,\mathbb V)$ is defined in \eqref{wie5l}.

If $\xi$ is the velocity vector field of the flow $\{\tau_t
\}_{t\in\mathbb R}$, then one can consider $-{\cal L}_\xi$ as the
generating operator for $\{\Sigma_t \}_{t\in\mathbb R}$. On the
other hand the generator $Q_{(F,X)}$ of the flow
$\{\widetilde{\Sigma}_t \}_{t\in\mathbb R}$
has the form
\begin{equation}\label{osi4e}
    Q_{(F,X)}:=-(\nabla_X+\widetilde{F}),
\end{equation}
where $(F,X)\in {\cal F}_G^0$. The operator $\nabla_X$ is the
covariant derivative with respect to vector field $X\in
C^\infty(M,TM)$ and the second ingredient $\widetilde{F}$ of the
right-hand-side of (\ref{osi4e}) is an endomorphism of
$C^\infty(M,\mathbb{V})$ defined by the function $F\in pr_1({\cal
F}_G^0)$ in the following way
\begin{equation}\label{chw48os}
    \widetilde{F}([(p,v)]):=[(p,F(p)v)].
\end{equation}
From (\ref{sp98la}) we find that
\begin{equation}\label{sp98laer}
{\cal R}\circ {\cal L}_\xi=(\nabla_X+\widetilde{F})  \circ{\cal R}.
\end{equation}
Since $[{\cal L}_\xi,{\cal L}_\eta]={\cal L}_{[\xi,\eta]}$ we see
from \eqref{chw48os} that the linear monomorphism
\begin{equation}\label{ku7uk}
    Q:{\cal F}_G^0\longrightarrow\mbox{End}(C^\infty(M,\mathbb{V}))
\end{equation}
satisfies the prequantization property
\begin{equation}\label{pre32q1}
    [Q_{(F,X)},Q_{(G,Y)}]=Q_{\{\!\!\{(F,X),(G,Y)\}\!\!\}},
\end{equation}
where the bracket $[\cdot,\cdot]$ on the left-hand-side of
(\ref{pre32q1}) is the commutator of the 1-st order differential
operators and the Lie bracket $\{\!\!\{\cdot,\cdot\}\!\!\}$ is
defined in (\ref{lie423spr}). In the non-degenerate case, i.e. when
$(F,X)$ is defined by $F$ (see (\ref{h71am87i})) the property
(\ref{pre32q1}) reduces to
\begin{equation}\label{pre32q13}
    [Q_F,Q_G]=Q_{\{F,G\}},
\end{equation}
where $Q_F:=Q_{(F,X_F)}$ and the bracket $\{F,G\}$ is defined by
\begin{equation}\label{pr9kl}
\{F,G\}:=-2\Omega(X^*_F,Y^*_G)+[F,G].
\end{equation}

In the case $G=U(1)$ the operator (\ref{osi4e}) is the
Kostant--Souriau prequantization operator. So, one can consider the
construction presented above as a natural generalization of the
Kostant--Souriau prequantization procedure.

\section {Quantization}

It is well known that in order to quantize a function $F\in
C^\infty_{U(1)}(P,i\mathbb R)\cong C^\infty(M,\mathbb R)$ (a
classical physical quantity) in the Kostant--Souriau geometric
quantization one needs to choose a proper polarization ${\cal
P}\subset T^{\mathbb C}M$ on the symplectic manifold $(M,\Omega)$.
Further, using ${\cal P}$ one realizes Hilbert space ${\cal H}$ by
sections of the vector bundle $\widetilde{\pi}:\mathbb{V}\rightarrow
M$ in such a way that differential operator $Q_F$, defined in
(\ref{osi4e}), preserves ${\cal H}$ and admits a self-adjoint
extension in it, (for details see e.g. \cite{S}).

In the method of quantization which will be discussed here we avoid
the notion of polarization. For the construction of the Hilbert
space ${\cal H}$ we will instead use the ${\cal B}(V)$-valued
positive definite kernel discussed in Section 2. In fact Hilbert
space $\mathcal K\cong{\cal H}$ was defined in Section 2 as  one
of the triple of equivalent objects used for the description of
${\cal B}(V)$-valued positive definite kernels.

In Section 3 we described the flows $\{\tau_t \}_{t\in\mathbb
R}\subset \mbox{Aut}(P,\vartheta)$ of automorphisms of the principal
bundle $\pi:P\rightarrow M$ with the fixed
$DT(e)(\mathfrak{g})$-valued connection form $\vartheta$, see
(\ref{je3sze}). Here we restrict ourselves to those flows
$\{\tau_t \}_{t\in\mathbb R}\subset\mbox{Aut}(P,K)\subset
\mbox{Aut}(P,\vartheta)$ which preserve ${\cal
B}(V)$-valued positive definite kernel $K$, i.e. such ones that
\begin{equation}\label{ewgfr2q}
    K(\tau_t(p),\tau_t(q))=K(p,q),
\end{equation}
for any $p,q\in P$ and $t\in\mathbb{R}$.

The following statement is valid
\begin{theorem}\label{th:4.1}
The flow $\{\tau_t \}_{t\in\mathbb R}\subset\mbox{Aut}(P)$ satisfies
invariance condition (\ref{ewgfr2q}) if and only if there exists a
unitary flow $U_t:{\cal H}\rightarrow{\cal H}$  such that
\begin{equation}\label{g5r43}
    \mathfrak{K}(\tau_t(p))=U_t\mathfrak{K}(p),
\end{equation}
where $\mathfrak{K}:P\rightarrow \mathcal{B}(V,{\cal H})$ is the map
satisfying the condition (\ref{gyt5}) of the definition
\textbf{(ii)} and is related to the kernel $K(p,q)$ by
(\ref{f4r54}).
\end{theorem}{\it Proof}:

Provided the map $\mathfrak{K}:P\rightarrow \mathcal{B}(V,{\cal H})$
has property (\ref{g5r43}) we obtain (\ref{ewgfr2q}) from the
equality (\ref{f4r54}).

Let us take $f,g\in {\cal K}_0$, where elements of the vector
subspace ${\cal K}_0$ are defined in (\ref{Afr432}). We define the
flow $\{U_t \}_{t\in\mathbb R}$ on ${\cal K}_0$ by
\begin{equation}\label{fl81q}
    (U_tf)(p):=f(\tau_{-t}(p))=\sum_{i=1}^I K(\tau_{-t}(p),p_i)v_i.
\end{equation}
The invariance condition (\ref{ewgfr2q}) and (\ref{Afr432}) implies
the equality
\begin{equation}\label{uni4321}
    \langle U_tf|U_tg\rangle=\langle f|g\rangle
\end{equation}
for the scalar product $\langle \cdot|\cdot\rangle$ defined in
(\ref{sc3e}). So, we can consider $\{U_t \}_{t\in\mathbb R}$ us a
unitary flow on the Hilbert space ${\cal H}:=\overline{{\cal K}_0}$.
Let $E_p:{\cal H}\rightarrow V$ be the evaluation functional at
$p\in P$. Let us define the map $\mathfrak{K}:P\rightarrow
\mathcal{B}(V,{\cal H})$ by $\mathfrak{K}(p):=E_p^*$. Then for any
$v\in V$ and $f\in{\cal K}_0$ one has
\begin{equation}\label{os32tat}
\langle\mathfrak{K}(\tau_t(p))v |f \rangle=\langle E^*_{\tau_t(p)}v
|f \rangle=\langle v , E_{\tau_t(p)}f \rangle=\langle v,\sum_{i=1}^I
K(\tau_{t}(p),p_i)v_i\rangle=
\end{equation}
\begin{equation}\label{os32tatde}
=\langle v, E_p U_{-t}f\rangle=\langle\mathfrak{K}(p)v |U_{-t}f
\rangle = \langle U_{t}\mathfrak{K}(p)v |f \rangle.
\end{equation}
Thus we obtain (\ref{g5r43}).

\hfill $\Box$

The above theorem implies
\begin{proposition}\label{prop:4.2}
For any flow $\{\tau_t \}_{t\in\mathbb R}\subset\mbox{Aut}(P,K)$ one
has
\begin{equation}\label{g5r43fre}
    \bfig \square/>`>`>`>/<700,500>[ \mathbb{V}`{\cal
H}`\mathbb{V}`{\cal
H};\widetilde{\mathfrak{K}}`\widetilde{\tau}_t`U_t`\widetilde{\mathfrak{K}}]
\efig,
\end{equation}
i.e.
$U_t\circ\widetilde{\mathfrak{K}}=\widetilde{\mathfrak{K}}
\circ\widetilde{\tau}_t$
for $t\in\mathbb{R}$, where $\widetilde{\mathfrak{K}}$ is the
coherent state map defined in (\ref{desw239}) and the flow
$\widetilde{\tau}_t:\mathbb{V}\rightarrow\mathbb{V}$ is given in
(\ref{wi8we}). Moreover, if $K$ is a smooth positive definite
kernel, then ${\cal H}_0:={\mbox
span}\{\widetilde{\mathfrak{K}}(\mathbb{V})\}\subset{\cal H}$ is an
essential domain of the generator $\widehat{F}$ of the flow
$U_t=:e^{i\widehat{F}t}$.

\end{proposition}{\it Proof}:

The equivariance property (\ref{g5r43fre}) follows from the
definition (\ref{desw239}) and the property (\ref{g5r43}).

From (\ref{g5r43}) we have
\begin{equation}\label{f4e1q}
e^{i\widehat{F}t}\widetilde{\mathfrak{K}}([(p,v)])=
\widetilde{\mathfrak{K}}(\widetilde{\tau}_t([(p,v)])).
\end{equation}
Since the coherent state map $\widetilde{\mathfrak{K}}$ is smooth
(see Proposition \ref{glad3}) and $\{{\widetilde{\tau}}_t\}_{t\in
\mathbb{R}}$ is a smooth flow we find that the right-hand-side of
(\ref{f4e1q}) is differentiable with respect to $t$. Then by Stone
theorem $\widetilde{\mathfrak{K}}([(p,v)])\in {\cal
D}(\widehat{F})$, i.e. ${\cal H}_{0}\subset {\cal D}(\widehat{F})$.
Moreover, the vector subspace ${\cal H}_{0}\subset{\cal H}$ is
invariant with respect to the action of $e^{i\widehat{F}t}$. Thus,
since ${\cal H}_{0}$ is dense in ${\cal H}$, it turns to be (see
\cite{R-S}, Volume 1, Theorem VIII.11) an essential domain of
$\widehat{F}$, i.e.
\begin{equation}\label{e3dq}
    \overline{\widehat{F}_{|{\cal H}_{0}}}=\widehat{F},
\end{equation}
where the bar in (\ref{e3dq}) denotes the closure of the symmetric
operator $\widehat{F}_{|{\cal H}_{0}}$.

\hfill $\Box$

Note that within the representation of Hilbert space ${\cal H}$ by
the $V$-valued functions on $P$, see (\ref{ne3e}), one has
\begin{equation}\label{jot5q}
    U_t=J^{-1}\circ\Sigma_t\circ J.
\end{equation}
Thus for generating operator $\widehat{F}$ we obtain
\begin{equation}\label{gfjot5q}
    \widehat{F}=iJ^{-1}\circ{\cal L}_\xi\circ J.
\end{equation}
Taking realization of ${\cal H}$ by sections of the bundle
$\widetilde{\pi}:\mathbb{V}\rightarrow M$, see (\ref{wie5l}) and
(\ref{osi4e}), one has
\begin{equation}\label{it5q}
    U_t=I^{-1}\circ\widetilde{\Sigma}_t\circ I
\end{equation}
and
\begin{equation}\label{32it5q}
    \widehat{F}=iI^{-1}\circ(\nabla_X+\widetilde{F})\circ I.
\end{equation}
Let us recall in this context that $J({\cal H}_0)={\cal K}_0$. Note,
also that the above two representations of $\{U_t \}_{t\in\mathbb
R}$ are intertwined by the operator ${\cal R}$ defined in
(\ref{er2re}).

\vspace{2ex}

Since ${\cal H}_{0}$ is an essential domain of $\widehat{F}$ we have
the commutation relation
\begin{equation}\label{sp8tqrt}
\widehat{F}e^{i\widehat{F}t} =e^{i\widehat{F}t}\widehat{F},
\end{equation}
valid on the elements of ${\cal H}_0$. Now, let us define the vector
subspace ${\cal U}_1:={\cal H}_{0}+\widehat{F}({\cal H}_{0})\subset
{\cal H}$. From (\ref{sp8tqrt}) we obtain, in particular, that
$e^{i\widehat{F}t}{\cal U}_{1}\subset {\cal U}_{1}$. Due to
(\ref{gfjot5q}) one obtains that for a smooth vector field $\xi$ and
smooth coherent state map $\widetilde{\mathfrak{K}}$, the
left-hand-side of
$\widehat{F}e^{i\widehat{F}t}\widetilde{\mathfrak{K}}([(p,v)])
=e^{i\widehat{F}t}\widehat{F}\widetilde{\mathfrak{K}}([(p,v)])$ is
differentiable with respect to $t$. Then by Stone theorem
$\widehat{F}\widetilde{\mathfrak{K}}([(p,v)])\in{\cal
D}(\widehat{F})$ and $\widetilde{\mathfrak{K}}([(p,v)])\in{\cal
D}(\widehat{F}^2)$. Thus we have ${\cal U}_{1}\subset{\cal
D}(\widehat{F})$ and ${\cal H}_{0}\subset {\cal D}(\widehat{F}^2)$.
In such a way, step by step, we prove that ${\cal U}_l\subset{\cal
D}(\widehat{F})$, where ${\cal U}_l$ is defined by
 \begin{equation} \label{tre99aq}
{\cal U}_{l}:=  {\cal U}_{l-1} + \widehat{F} ({\cal
U}_{l-1}),\;\;\;{\cal U}_{0}:={\cal H}_{0},
\end{equation}
for $l=1,2,\ldots$. By ${\cal U}_{\infty}$ we denote the vector
space spanned by all ${\cal U}_{l}$, $l\in\mathbb{N}\cup\{0\}$.
Summing up the above considerations we formulate the following
\begin{proposition}
One has the filtration
\begin{equation}\label{filtrq}
{\cal U}_{0}\subset{\cal U}_{1}\subset\ldots\subset{\cal U}_{\infty}
\subset{\cal D}(\widehat{F})
\end{equation}
of the domain ${\cal D}(\widehat{F})$ of the operator $\widehat{F}$
onto its essential domains. This filtration is preserved
\begin{equation}\label{bbb99a5q}
e^{i\widehat{F}t}{\cal U}_{l}\subset {\cal U}_{l},
\end{equation}
by the flow $\{e^{i\widehat{F}t} \}_{t\in\mathbb R}$. Moreover
\begin{equation}\label{cccyclq}
\widehat{F}{\cal U}_{l}\subset {\cal U}_{l+1}
\end{equation}
and
\begin{equation}\label{bbb99a2q}
{\cal U}_{\infty}\subset {\cal D}(\widehat{F}^l),
\end{equation}
for $l\in\mathbb{N}\cup\{0\}$.
\end{proposition}

Next proposition shows how to reconstract the classical Hamiltonian
$F$ from the quantum Hamiltonian
$\hat F$.
\begin{proposition}
The generating function $F:P\to DT(e)(\mathfrak{g})$ is obtained as
the coherent states mean values function of $\hat F$, i.e.
\begin{equation}
F(p)=i(\mathfrak K(p)^*\mathfrak K(p))^{-1}\mathfrak K^*(p)\hat F
\mathfrak K(p).
\end{equation}
\end{proposition}{\it Proof}:

Using $U_t=e^{it\hat F}$, from \eqref{je3sze}, \eqref{roz312w},
and \eqref{g5r43} we have
\begin{equation}
\mathfrak K(p)^*\hat F\mathfrak K(p)=
-i\frac{d}{dt}[\mathfrak K^*(p)U_t\mathfrak K(p)]_{|t=0}=-i\mathfrak K^*(p)
\frac{d}{dt}\mathfrak K(\tau_t(p))_{|t=0}=
\end{equation}\begin{equation}
=-i\mathfrak K^*(p)(\xi \mathfrak K)(p)=
-i\mathfrak K^*(p)\mathfrak K(p)\vartheta(\xi)(p)
\end{equation}\begin{equation}
=-i\mathfrak K^*(p)\mathfrak K(p)\vartheta(F^*)(p)=
-i\mathfrak K^*(p)\mathfrak K(p)F(p)
\end{equation}

\hfill $\Box$

\section{The coordinate description and examples}

In this section we will investigate the quantization procedure
which was proposed in Section 4 in terms of a concrete trivialization
of the principal bundle $\pi: P\to M$. Because of its importance
for physical application we will discuss the holomorphic case in details.
Finally we will present two examples, where in the second example
we obtain a holomorphic (anti--holomorphic) realization for any
self--adjoint operator with simple spectrum.

Now, for the further investigation of the quantum flow generator
$\widehat{F}$ given in (\ref{32it5q}) we will describe its
representation in a trivialization
\begin{equation}\label{tri54c}
    s_\alpha:\Omega_\alpha\rightarrow P,\;\;\;\pi\circ
    s_\alpha=id_{\Omega_\alpha}
\end{equation}
 of $\pi: P\rightarrow M$,
where $\bigcup_{\alpha\in A}\Omega_\alpha=M$ is a covering of $M$ by
the open subsets.

We note that on $\pi^{-1}(\Omega_\alpha)$ one has
\begin{equation}\label{lok8kr}
\Omega(p)=T(h^{-1})\left(d\vartheta_\alpha(m)+\frac{1}{2}[\vartheta_\alpha(m),
\vartheta_\alpha(m)]\right)T(h),
\end{equation}
\begin{equation}\label{lok8r}
\textbf{D}F(p)=T(h^{-1})\left(dF_\alpha(m)+[\vartheta_\alpha(m),
F_\alpha(m)]\right)T(h),
\end{equation}
for $p=s_\alpha(m)h$,
where
\begin{equation}\label{fr56tg}
\vartheta_\alpha:=s_\alpha^*\vartheta\;\;\;and\;\;\;F_\alpha:=F\circ
s_\alpha.
\end{equation}

The positive definite kernel $K:P\times P\rightarrow{\cal B}(V)$ in
the trivialization (\ref{tri54c}) is described by
\begin{equation}\label{lok8j}
    \mathfrak{K}_\alpha(m):=\mathfrak{K}\circ s_\alpha(m),
\end{equation}
\begin{equation}
K_{\overline{\alpha}\beta}(m,n):=
\mathfrak{K}^*_\alpha(m)\mathfrak{K}_\beta(n),
\end{equation}
for $m\in\Omega_\alpha$ and $n\in\Omega_\beta$. Using (\ref{fr56tg})
from (\ref{je3sze}) we obtain
\begin{equation}\label{kon40baz}
\vartheta_\alpha(m)=\left(\mathfrak{K}_\alpha(m)^*\mathfrak{K}_\alpha(m)
\right)^{-1}\mathfrak{K}_\alpha(m)^*d\mathfrak{K}_\alpha(m).
\end{equation}

Let us take the flow $\{\tau_t \}_{t\in\mathbb
R}\subset\mbox{Aut}(P,K)$ and define local cocycle
${]-\varepsilon,\varepsilon[\times} \Omega_\alpha \ni (t,m)\mapsto
g_\alpha(t,m)\in G$
by
\begin{equation}\label{spr554o}
    \tau_t(s_\alpha(m))=s_\alpha(\sigma_t(m))g_\alpha(t,m).
\end{equation}
For $p=s_\alpha(m)h$ we have
\begin{equation}\label{gru77u}
\mathfrak{K}(\tau_t(p))=\mathfrak{K}_\alpha(\sigma_t(m))T(g_\alpha(t,m))T(h).
\end{equation}
From (\ref{gru77u}) and (\ref{g5r43}) one gets
\begin{equation}\label{op676roz}
i\widehat{F}\mathfrak{K}_\alpha(m)v=(X\mathfrak{K}_\alpha)(m)v+
\mathfrak{K}_\alpha(m)\phi_\alpha(m)v,
\end{equation}
where $v\in V$ and
\begin{equation}
\phi_\alpha(m):=\frac{d}{dt}T(g_\alpha(t,m))_{|t=0}.
\end{equation}
\begin{proposition}
If we assume that $\xi=X^*-F^*\in{\cal E}_G^0$ is the velocity
vector field for $\{\tau_t \}_{t\in\mathbb R}$, then the map
$\phi_\alpha:\Omega_\alpha\rightarrow{\cal B}(V)$ is given by
\begin{equation}\label{cztery656}
-\phi_\alpha=F_\alpha+\vartheta_\alpha(X)
\end{equation}
\end{proposition}{\it Proof}:

From (\ref{int6r}), (\ref{spr554o}) and
\begin{equation}\label{zm78ec}
    \tau_t^h(s_\alpha(m))=s_\alpha(\sigma_t(m))\kappa_\alpha(t,m)
\end{equation}
we find that
\begin{equation}\label{d345e67}
\widetilde{\tau}_t[(s_\alpha(m),v)]=[(s_\alpha(\sigma_t(m)),
T(g_\alpha(t,m)^{-1})v]
\end{equation}
and
\begin{equation}\label{d345e45}
\widetilde{\tau}_t[(s_\alpha(m),v)]=[(s_\alpha(\sigma_t(m)),
T((\kappa_\alpha(t,m)c(t,s_\alpha(m)))^{-1})v].
\end{equation}
Comparing (\ref{d345e67}) and (\ref{d345e45}) one obtains
\begin{equation}\label{d3f5e67}
T(g_\alpha(t,m))=T(\kappa_\alpha(t,m)c(t,s_\alpha(m))).
\end{equation}
Differentiating (\ref{d3f5e67}) at $t=0$ and taking into account
\begin{equation}
\vartheta(s_\alpha(m)h)=T(h^{-1})\vartheta_\alpha(m)T(h)+T(h^{-1})dT(h),
\end{equation}
where $h\in G$, gives (\ref{cztery656}).

\hfill $\Box$

Using (\ref{lok8kr}) and (\ref{lok8r}) we find that
$\xi=X^*-F^*\in{\cal E}_G^0$, i.e, ${\cal L}_\xi\vartheta=0$, if and
only if
\begin{equation}\label{duz42wi}
{\cal L}_X\vartheta_\alpha\equiv X\llcorner
d\vartheta_\alpha+d(\vartheta_\alpha(X))=d\phi_\alpha+[\vartheta_\alpha,
\phi_\alpha].
\end{equation}

The selfadjointess of $\widehat{F}$ implies the following relation
\begin{equation}\label{samo81s}
\mathfrak{K}_\beta(n)^*(X\mathfrak{K}_\alpha)(m)+
(X\mathfrak{K}_\beta)(n)^*\mathfrak{K}_\alpha(m)+
\mathfrak{K}_\beta(n)^*\mathfrak{K}_\alpha(m)\phi_\alpha(m)+
\phi_\beta(n)^*\mathfrak{K}_\beta(n)^*\mathfrak{K}_\alpha(m)\equiv0
\end{equation}
between the kernel map
$\mathfrak{K}_\alpha:\Omega_\alpha\rightarrow{\cal B}(V,{\cal H})$
and $(F,X)\in{\cal F}_G^0$.

The transition cocycle
$g_{\alpha\beta}:\Omega_\alpha\cap\Omega_\beta\rightarrow G$ defined
by
\begin{equation}\label{oczy776}
    s_\beta(m)=s_\alpha(m)g_{\alpha\beta}(m),
\end{equation}
for $m\in\Omega_\alpha\cap\Omega_\beta$ leads to the corresponding
gauge transformation of the formulae (\ref{op676roz}) which is given
by
\begin{equation}\label{tr5koh}
\mathfrak{K}_\beta(m)=\mathfrak{K}_\alpha(m)T(g_{\alpha\beta}(m))
\end{equation}
and
\begin{equation}\label{tr54fi}
\phi_\beta(m)=T(g_{\beta\alpha}(m))\phi_\alpha(m)T(g_{\alpha\beta}(m))+
\frac{d}{dt}T(g_{\beta\alpha}(\sigma_t(m))_{|t=0}T(g_{\alpha\beta}(m)).
\end{equation}

For the sake of completeness of our exposition in a fixed gauge let
us find the expression for the action of Kostant--Souriau operator
$Q_{(F,X)}=i I\circ \hat F\circ I^{-1}$ on the part of its
essential domain spanned by sections of the form $I(\psi)$, where
$\psi=\mathfrak K_\beta(n)v$ for $n\in\Omega_\beta$, $v\in V$.

In the $s_\alpha$-gauge section $I(\psi)\in C^\infty(M,\mathbb V)$ and
$Q_{(F,X)}I(\psi)$ are given by
\begin{equation}I(\psi)(m)=
[(s_\alpha(m),\mathfrak K^*_\alpha(m)\mathfrak K_\beta(n)v)]
\end{equation}
and by
\begin{equation}
(Q_{(F,X)}I(\psi))(m)=iI(\hat F\psi)(m)=[(s_\alpha(m),i\mathfrak
K^*_\alpha(m)\hat F \mathfrak K_\beta(n)v)]\end{equation}
respectively, $m\in\Omega_\alpha$. Hence, using the relation
\eqref{samo81s} we obtain the coordinate expression on $Q_{(F,X)}$
in terms of the kernel $K_{\bar\alpha\beta}(m,n)$:
\begin{equation}
Q_{(F,X)}(K_{\bar\alpha\beta}(\cdot,n))(m)v=
-(XK_{\bar\alpha\beta}(\cdot,n))(m)v-\phi_\alpha(m)^*K_{\bar\alpha\beta}(m,n)v.
\end{equation}

Recall here that the operator-valued maps
$\phi_\alpha:\Omega_\alpha\to\mathcal B(V)$ are related to the
generating function $F$ by \eqref{cztery656}.

From the view point of physical applications, see e.g.
\cite{H-O,O-S,H-O-T}, one of the most interesting cases appears when
$\pi:P\to M$ is a complex analytic principal $GL(N,\mathbb
C)$-bundle. Consequently we will assume that the coherent state map
$\mathfrak K:P\to \mathcal B(V,\mathcal H)$ is a complex analytic
map which satisfies the condition \eqref{d443e} for $T=id$ and $g\in
GL(N,\mathbb C)\cong GL(V,\mathbb C)$. Taking a complex analytic
trivialization $s_\alpha^{hol}:\Omega_\alpha\to P$ we find that
$\mathfrak K_\alpha^{hol}=\mathfrak K\circ
s_\alpha^{hol}:\Omega_\alpha\to \mathcal B(V,\mathcal H)$ is
holomorphic map and so does the transition map
$h_{\alpha\beta}:\Omega_\alpha\cap \Omega_\beta\to GL(N,\mathbb C)$,
where $s_\alpha^{hol}(m)=s_\beta^{hol}(m)h_{\alpha\beta}(m)$. Thus
the kernel $K^{hol}_{\bar\alpha\beta}= (\mathfrak
K^{hol}_\alpha)^*\mathfrak K^{hol}_{\beta}:
\Omega_\alpha\times\Omega_\beta\to GL(N,\mathbb C)$ is a map
anti-holomorphic in the first argument and holomorphic in the second
one. Using $K^{hol}_{\bar\alpha\beta}$ we define another
trivialization
\begin{equation}\label{triv432}
s_\alpha(m):=s_\alpha^{hol}(m)K^{hol}_{\bar
\alpha\alpha}(m,m)^{-\frac12}
\end{equation}
with a transition cocycle
\begin{equation}
 g_{\alpha\beta}(m):=K^{hol}_{\bar \alpha\alpha}(m,m)^{\frac12}h_{\alpha\beta}(m)
K^{hol}_{\bar \beta\beta}(m,m)^{-\frac12}.
\end{equation}
It follows from $\mathfrak K^{hol}_\beta(m)=\mathfrak
K_\alpha^{hol}(m)h_{\alpha\beta}(m)$ that
$g_{\alpha\beta}:\Omega_\alpha\cup\Omega_\beta\to U(N)\subset
GL(N,\mathbb C)$. So, we can reduce the holomorphic coherent state
map $\mathfrak K:P\to \mathcal B(V,\mathcal H)$ to the principal
$U(N)$-bundle $\pi:P^u\to M$ which is a subbundle of $\pi:P\to M$
defined by the trivialization \eqref{triv432}. Therefore, we can
apply the method of quantization investigated in this section to the
holomorphic case.

In this connection we note that holomorphic flow
$\{\tau_t\}_{t\in\mathbb R}\subset Aut(P)$ preserves the kernel
$K^{hol}$ defined by $K^{hol}_{\bar\alpha\beta}$ if and only if
\begin{equation}
K^{hol}_{\bar\alpha\beta}(\sigma_t (m),\sigma_t (n))=
h_\alpha(t,m)^*K^{hol}_{\bar\alpha\beta}(m,n)h_\beta(t,n),
\end{equation}
where the holomorphic cocycle $h_\alpha(t,m)$ is defined by
\begin{equation}\label{coc324}
 \tau_t(s_\alpha^{hol}(m))=s_\alpha^{hol}(\sigma_t(m))h_\alpha(t,m)
\end{equation}
The cocycles $g_\alpha(t,m)$ and $h_\alpha(t,m)$ corresponding to
the sections $s_\alpha:\Omega_\alpha\to P$ and
$s_\alpha^{hol}:\Omega_\alpha\to P$, respectively, are related by
\begin{equation}\label{cztczto}
 g_\alpha(t,m)=K^{hol}_{\bar\alpha\alpha}(\sigma_t(m),\sigma_t(m))^{\frac12}
 h_\alpha(t,m)
K^{hol}_{\bar\alpha\alpha}(m,m)^{-\frac12},
\end{equation}
where $g_\alpha(t,m)$ is defined in (\ref{spr554o}).

Note here that
\begin{equation}\label{53hgfs}
\mathfrak{K}_\alpha(m)=
\mathfrak{K}^{hol}_\alpha(m)K^{hol}_{\bar\alpha\alpha}(m,m)^{-\frac12}.
\end{equation}

Using \eqref{cztczto} we find that
\begin{equation}\label{phd421}
 \phi_\alpha(m)=K^{hol}_{\bar\alpha\alpha}(m,m)^{\frac12}
 \phi_\alpha^{hol}(m)K^{hol}_{\bar\alpha\alpha}(m,m)^{-\frac12}-
 K^{hol}_{\bar\alpha\alpha}(m,m)^{\frac12}\frac{d}{dt}K^{hol}_{\bar\alpha\alpha}
 (\sigma_t(m),\sigma_t(m))^{-\frac12}|_{t=0},
\end{equation}
where $\phi_\alpha^{hol}:\Omega_\alpha\to GL(N,\mathbb C)$ defined
by
\begin{equation}
 \phi^{hol}_\alpha(m):=\frac{d}{dt}h_\alpha(t,m)|_{t=0}
\end{equation}
is a holomorphic map.

One also has
\begin{equation}
 F_\alpha(m)=K^{hol}_{\bar\alpha\alpha}(m,m)^{\frac12}F_\alpha^{hol}(m)
 K^{hol}_{\bar\alpha\alpha}(m,m)^{-\frac12}
\end{equation}
where $F_\alpha=F\circ s_\alpha$ and $F_\alpha^{hol}=F\circ
s_\alpha^{hol}$ and
\begin{equation}
 \vartheta_\alpha(m)=
 K^{hol}_{\bar\alpha\alpha}(m,m)^{\frac12}\vartheta_\alpha^{hol}(m)
 K^{hol}_{\bar\alpha\alpha}(m,m)^{-\frac12}+
 K^{hol}_{\bar\alpha\alpha}(m,m)^{\frac12}dK^{hol}_{\bar\alpha\alpha}
 (m,m)^{-\frac12}.
\end{equation}

From \eqref{53hgfs} and \eqref{phd421} we obtain the holomorphic representation
\begin{equation}\label{hyy432}
 i\hat F\mathfrak K_\alpha^{hol}(m)v=(X^{(1,0)}\mathfrak K^{hol}_\alpha)(m)v+
 \mathfrak K_\alpha^{hol}(m)\phi_\alpha^{hol}(m)v
\end{equation}
of the generating operator $\hat F$. Vector field $X^{(1,0)}$
appearing in \eqref{hyy432} is $(1,0)$-component of the vector field
$X=X^{(1,0)}+X^{(0,1)}$ tangent to the flow $\{\tau_t\}_{t\in\mathbb
R}$. Note that $X^{(0,1)}=\overline{ X^{(1,0)}}$ and vectors
$\mathfrak K_\alpha^{hol}(m)v\in\mathcal H$, such that
$m\in\Omega_\alpha$, $v\in V$ and $\alpha\in A$, span an essential
domain of $\hat F$.

Using \eqref{hyy432} we obtain anti-holomorphic representation of
Kostant-Souriau operator
\begin{equation}\label{j2332j}
 (Q_{(F,X)}K^{hol}_{\bar\alpha\beta}(\cdot,n))(m)v=
 -(X^{(0,1)}K^{hol}_{\bar\alpha\beta}(\cdot,n))(m)v-
 \phi_\alpha^{hol}(m)^*K_{\bar\alpha\beta}(m,n)v.
\end{equation}

Hence we see, that in the holomorphic case the essential domain of
$Q_{(F,X)}$ consists of those  anti-holomorphic sections of
$\pi:\mathbb V\to M$ which are locally spanned by
$K_{\bar\alpha\beta}^{hol}(\cdot,n)v$.

Finally we present two simple examples illustrating our method of
quantization.

\paragraph*{Example 1} We consider the trivial principal
$GL(2,\mathbb C)$-bundle $P=M\times GL(2,\mathbb C)$ where
$M=\mathbb D\times \mathbb D$ is the product of two unit discs
$\mathbb D\subset \mathbb C$. Let $z=\left(z_1 \atop z_2\right)$,
$w=\left(w_1 \atop w_2\right)$ be elements of $\mathbb D\times
\mathbb D$. We introduce the positive definite kernel
$K^{hol}:P\times P\to \mathcal B(\mathbb C^2)$ which in the standard
trivialization $s(z):=(z,\left(1\;0 \atop{0\;1}\right))$ has the
form
\begin{equation}\label{weqs1}
 K_{\mathbb D\times \mathbb D}^{hol}(\bar z,w):=\begin{pmatrix}
1+\frac{1}{1-\bar z_1w_1}& w_1\\
\bar z_1& 1+\frac{1}{1-\bar z_2w_2}+\bar z_1w_1
                                          \end{pmatrix}
\end{equation}
and the holomorphic flow $\sigma_t:\mathbb D\times \mathbb D\to \mathbb
D\times \mathbb D$, $t\in\mathbb R$, defined by
\begin{equation}\label{asdads}
 \sigma_t\left( z_1 \atop z_2 \right):=\left( e^{it}z_1 \atop {e^{-it}z_2}
 \right).
\end{equation}
Kernel \eqref{weqs1} satisfies the relationship
\begin{equation}
 K^{hol}_{\mathbb D\times \mathbb D}(\overline{\sigma_t(z)},\sigma_t (w))=
 h^{+}(t)K^{hol}_{\mathbb D\times \mathbb D}(\bar z,w)h(t),
\end{equation}
where $h(t)=\left(e^{it}\;\;\;\;\;0 \atop{0\;\;\;\;e^{-it}}\right)$,
$t\in\mathbb R$. Thus the flow $\tau_t:P\to P$ defined by $\tau_t
(s(z))=s(\sigma_t (z))h(t)$ preserves the positive definite kernel
\begin{equation}
 K^{hol}\big((\bar z,g^+),(w,h)\big):=
 g^{+}K^{hol}_{\mathbb D\times \mathbb D}(\bar z,w)h,
\end{equation}
where $(z,g),(w,h)\in (\mathbb D\times \mathbb D)\times GL(2,\mathbb
C)$.

The vector field
\begin{equation}
 X=i\left(z_1\frac{\partial}{\partial z_1}+z_2\frac{\partial}{\partial z_2}-
\bar z_1\frac{\partial}{\partial \bar z_1}-
\bar z_2\frac{\partial}{\partial \bar z_2}\right)
\end{equation}
tangent to the one-parameter group \eqref{asdads} and function
$F:\mathbb D\times\mathbb D\times U(2)\to Mat_{2\times2}(\mathbb C)$
defined by
\begin{equation}
F(z,g)=g^{-1}K^{hol}(\bar z,z)^{\frac12}F^{hol}(z)
 K^{hol}(\bar z,z)^{-\frac12}g,
\end{equation}
where $F_\alpha^{hol}(m)$ is given by
\begin{equation}
F_\alpha^{hol}(m)=
\end{equation}
\begin{equation}
=\begin{pmatrix}
\frac{i(4-3|z_1|^2-2|z_2|^2+2|z_1|^4+|z_1|^2|z_2|^2-|z_1|^4|z_2|^2)}
{(1-|z_2|^2)(4-|z_1|^2-2|z_2|^2)}&
\frac{iz_1(1-|z_1|^2)(2-4|z_2|^2+|z_2|^4)}
{(1-|z_2|^2)(4-|z_1|^2-2|z_2|^2)}\\
\frac{i\bar z_1|z_1|^2(1-|z_2|^2)}
{(1-|z_1|^2)(4-|z_1|^2-2|z_2|^2)}&
\frac{i(1-|z_1|^2)(-4+2|z_1|^2+8|z_2|^2-2|z_2|^4-4|z_1|^2|z_2|^2-|z_1|^2|z_2|^4)}
{(1-|z_2|^2)(4-|z_1|^2-2|z_2|^2)}
                                          \end{pmatrix}\nonumber
\end{equation}
satisfy equation (\ref{h71am87i}). Applying the formulae
\eqref{j2332j} in this case we obtain Kostant--Souriau operator
\begin{equation}
(Q_{(F,X)}\psi)(\bar z_1,\bar z_2)=
i \begin{pmatrix} \bar z_1 \frac{\partial \psi_1(\bar z_1,\bar z_2)}
{\partial \bar z_1}+\psi_1(\bar z_1,\bar z_2)\\
\bar z_1 \frac{\partial \psi_2(\bar z_1,\bar z_2)}{\partial\bar
z_1}+\bar z_2\frac{\partial\psi_2(\bar z_1,\bar z_2)} {\partial \bar
z_2}-\psi_2(\bar z_1,\bar z_2)\end{pmatrix},
\end{equation}
where $\psi\in\mathcal D(Q_{(F,X)})$ is given by
\begin{equation}
\psi(\bar z_1,\bar z_2)=\begin{pmatrix}\psi_1(\bar z_1,\bar z_2)\\
\psi_2(\bar z_1,\bar z_2)\end{pmatrix}=
\begin{pmatrix}\sum_{k=1}^K \frac{v_{1k}}{1-\bar z_1 w_{1k}}+c \\
\sum_{k=1}^K \frac{v_{2k}}{1-\bar z_2 w_{2k}}+c  \bar z_1\end{pmatrix},
\end{equation}
and constant $c,v_{1k}$, $w_{1k}$, $v_{2k}$, $w_{2k}\in \mathbb{C}$
satisfy the condition
\begin{equation}
c=\sum_{k=1}^K v_{1k}+w_{1k}v_{2k}.
\end{equation}

The unitary flow generated by $Q_{(F,X)}$ is the quantization of the
flow (\ref{asdads}).

\paragraph*{Example 2}

Let $\hat F$ be a self-adjoint operator with simple spectrum acting
in Hilbert space $\mathcal H$. Fix standard Hilbert space
isomorphism $U:\mathcal H\to L^2(\mathbb R,d\sigma)$, where measure
$d\sigma$ is determined by spectral measure $dE$ of $\hat F$ and a
certain choice of normalized cyclic vector $|0\rangle\in\mathcal H$
for $\hat F$:
\begin{equation}
d\sigma(\omega):=\langle0|dE(\omega)|0\rangle,
\end{equation}
$\omega\in\mathcal R$. The homogeneous polynomials $\{\omega^n\}_{n=0}^\infty$
form a subset linearly dense in $L^2(\mathbb R,d\sigma)$. After Gram-Schmidt
orthonormalization procedure they give
an orthonormal polynomial basis $\{P_n\}_{n=0}^\infty$, $deg P_n=n$,
in $L^2(\mathbb R,d\sigma)$. Acting by
$P_n(\hat F)$ on $|0\rangle$ one obtains the orthonormal basis
\begin{equation}
|n\rangle :=P_n(\hat F)|0\rangle
\end{equation}
in Hilbert space $\mathcal H$.

We now assume the condition
\begin{equation}\label{momlim}
\limsup_{n\to\infty}\frac{\sqrt[n]{|\mu|_n}}{n}<+\infty
\end{equation}
on the absolute moments
\begin{equation}
|\mu|_n:=\int_{\mathbb R}|\omega|^n d\sigma(\omega)=
\frac{1}{P_0^2}\langle 0|\;|\hat F|^n\;|0\rangle
\end{equation}
of the operator $\hat F$. It follows from \eqref{momlim}
that there exists the maximal open strip $\Sigma\subset\mathbb C$
in complex plane $\mathbb C$, which is
invariant under the translations
\begin{equation}\label{trans5}
\sigma_t (z):=z+t
\end{equation}
$t\in\mathbb R$ and such that the characteristic functions
\begin{equation}
\chi(s)=\int_{\mathbb R} e^{-i\omega s} d\sigma(\omega),
\end{equation}
$s\in\mathbb R$, of the measure $d\sigma$ possesses holomorphic
extension $\chi_\Sigma$ to $\Sigma$, see \cite{H-O-T}.

Hence, one can define on the principal $U(1)$-bundle
$P:=\Sigma\times U(1)$ the positive definite kernel:
\begin{equation}
K[\overline{(z,g)},(v,h)]:=\bar g K_\Sigma(\bar z,v)h
\end{equation}
where
\begin{equation}\label{ker_sigma}
K_\Sigma(\bar z,v):=\chi_\Sigma(\bar z -v)
\end{equation}
and $(z,g),(v,h)\in\Sigma\times U(1)$. The map
$\mathfrak K_\Sigma:\Sigma\to\mathcal H\cong \mathcal B(\mathbb C,\mathcal H)$
defined by
\begin{equation}\label{k9}
\mathfrak K_\Sigma(\tau):=\sum_{n=0}^\infty \chi_n(z)|n\rangle
\end{equation}
where
\begin{equation}\label{chi10}
\chi_n(z):=\int e^{-iz\omega}P_n(\omega)d\sigma(\omega),
\end{equation}
for $z\in\Sigma$, gives factorization
\begin{equation}
K_\Sigma(\bar z,v)=\mathfrak K_\Sigma(z)^*\mathfrak K_\Sigma(v)
\end{equation}
of the kernel \eqref{ker_sigma}. From \eqref{k9} and \eqref{chi10} it follows that
\begin{equation}
e^{-it\hat F}\mathfrak K_\Sigma(z)=\mathfrak K_\Sigma(z+t).
\end{equation}
Thus we find that coherent states $\mathfrak K_\Sigma(z)$, $z\in
\Sigma$, span an essential domain $\mathcal D(\hat F)$ of $\hat F$
and
\begin{equation}
\hat F\mathfrak K_\Sigma(z)=i\frac{d}{dz}\mathfrak K_\Sigma(z).
\end{equation}
According to \eqref{lok8kr} the curvature form $\Omega_\Sigma$ of
the connection form $\vartheta_\Sigma$
defined by the kernel \eqref{k9} is given by
\begin{equation}
\Omega_\Sigma=i\partial\bar\partial(\log\circ K_\Sigma)(\bar z,z)=
i(\log\circ \chi_\Sigma)''(\bar z-z)d\bar z\wedge dz.
\end{equation}
For the vector field $X=\frac\partial{\partial z}+\frac\partial{\partial\bar z}$
tangent to the translation flow
\eqref{trans5} one has
\begin{equation}
X\llcorner\Omega_\Sigma=dF
\end{equation}
where
\begin{equation}\label{class16}
F=(\log\circ\chi_\Sigma)'(\bar z- z).
\end{equation}
We summarize the above facts in the proposition
\begin{proposition}
Operator $\hat F$ can be obtained by quantization of the classical
Hamiltonian $F$, given in \eqref{class16}, which generates
Hamiltonian flow \eqref{trans5} on symplectic manifold $(\Sigma,
\Omega_F)$.
 \end{proposition}

Representing $\hat F$ in the Hilbert space $I(\mathcal H)$ spanned by
the anti-holomorphic function
$K_\Sigma(\cdot,v)$, $v\in\Sigma$, we find for $F$ its Kostant--Souriau operator
\begin{equation}\label{KKS}
Q_F=I\circ\hat F\circ I^{-1}=i\frac{d}{d\bar z}.
\end{equation}
In conclusion let us note that our procedure of quantization,
applied to the case considered in this example, leads to realization
\eqref{KKS} of the operator $\hat F$ in function Hilbert space
$I(\mathcal H)$ which is an alternative to its spectral
representation in $L^2(\mathbb R,d\sigma)$.

\end{document}